\documentclass{aa}

\usepackage[varg]{txfonts}
\usepackage{siunitx}
\usepackage[version=4]{mhchem}
\usepackage{multirow}
\usepackage{graphicx}
\usepackage[caption=false]{subfig}
\usepackage[strict]{changepage}
\usepackage{xcolor}
\usepackage{ulem}
\usepackage[colorlinks=true,citecolor=blue]{hyperref}
\usepackage{float}
\usepackage{amssymb}
\usepackage{amsmath}
\usepackage{lscape}
\usepackage{mathtools}
\usepackage{natbib}
\usepackage{pdflscape}
\usepackage{gensymb}

\defcitealias{zhou2023}{Paper~I}   

\DeclareSIUnit\jansky{Jy}

\newcommand{\msun}{$\rm M_\odot$}

\newcommand{\ee}[1]{\mbox{${} \times 10^{#1}$}}

\newcolumntype{L}[1]{>{\raggedright\let\newline\\\arraybackslash\hspace{0pt}}m{#1}}
\newcolumntype{C}[1]{>{\centering\let\newline\\\arraybackslash\hspace{0pt}}m{#1}}
\newcolumntype{R}[1]{>{\raggedleft\let\newline\\\arraybackslash\hspace{0pt}}m{#1}}

\title{High-resolution APEX/LAsMA $^{12}$CO and $^{13}$CO (3-2) observation of the G333 giant molecular cloud complex : I. Evidence for gravitational acceleration in hub-filament systems}

\author{J. W. Zhou\inst{\ref{inst1}}
\and F. Wyrowski \inst{\ref{inst1}}
\and S. Neupane \inst{\ref{inst1}}
\and J. S. Urquhart \inst{\ref{inst2}}
\and N. J. Evans II \inst{\ref{inst3}}
\and E. Vázquez-Semadeni \inst{\ref{inst4}}
\and K. M. Menten
\inst{\ref{inst1}}
\and Y. Gong \inst{\ref{inst1}}
\and T. Liu 
\inst{\ref{inst5}}
}
\institute{Max-Planck-Institut f\"{u}r Radioastronomie, Auf dem H\"{u}gel 69, 53121 Bonn, Germany \label{inst1} \\
\email{jwzhou@mpifr-bonn.mpg.de}
\and
Centre for Astrophysics and Planetary Science, University of Kent, Canterbury, CT2 7NH, UK \label{inst2}
\and
Department of Astronomy, The University of Texas at Austin,
2515 Speedway, Stop C1400, Austin, Texas 78712-1205, USA \label{inst3}
\and
Instituto de Radioastronomía y Astrofísica, Universidad Nacional Autónoma de México, Antigua Carretera a Pátzcuaro \# 8701, Ex-Hda. San José de la Huerta, Morelia, Michoacán, México C.P. 58089 \label{inst4}
\and
Shanghai Astronomical Observatory, Chinese Academy of Sciences, 80 Nandan Road, Shanghai 200030, Peoples Republic of China \label{inst5} 
}
\date{Received xxx / Accepted xxx}

\abstract
{Hub-filament systems are suggested to be the birth cradles of high-mass stars and clusters.}
{We investigate the gas kinematics of hub-filament structures in the G333 giant molecular cloud complex using $^{13}$CO (3$-$2) observed with the APEX/LAsMA heterodyne camera.}
{We apply the FILFINDER algorithm to the integrated intensity maps of the $^{13}$CO $J$=3$-$2 line to identify filaments in the G333 complex, and we extract the velocity and intensity along the filament skeleton from moment maps. Clear velocity and density fluctuations are seen along the filaments, allowing us to fit velocity gradients around the intensity peaks.}
{The velocity gradients fitted to the LAsMA data and ALMA data agree with each other over the scales covered by ALMA observations in the ATOMS survey ($\textless$ 5 pc). Changes of velocity gradient with scale indicate a ''funnel'' structure of the velocity field in PPV space, indicative of a smooth, continuously increasing velocity gradient from large to small scales, and thus consistent with gravitational acceleration. The typical velocity gradient corresponding to a 1 pc scale is $\sim 1.6$ km s$^{-1}$ pc$^{-1}$. Assuming free-fall, we estimate a kinematic mass within 1 pc of $\sim$ 1190 M$_\odot$, which is consistent with typical masses of clumps in the ATLASGAL survey of massive clumps in the inner Galaxy. We find direct evidence for gravitational acceleration from comparison of the observed accelerations to those predicted by free-fall onto dense hubs with masses from millimeter continuum observations. On large scales, we find that the inflow may be driven by the larger scale structure, consistent with hierarchical structure in the molecular cloud and gas inflow from large to small scales. The hub-filament structures at different scales may be organized into a hierarchical system extending up to the largest scales probed, through the coupling of gravitational centers at different scales.}
{We argue that the ''funnel'' structure in PPV space can be an effective probe for the gravitational collapse motions in molecular clouds. 
The large scale gas inflow is driven by gravity, implying that the molecular clouds in G333 complex may be in the state of global gravitational collapse. 
}


\begin{document}

\titlerunning{Gas kinematics of multi-scale hub-filament structures}
\authorrunning{J.W. Zhou et al.}
\maketitle

\section{Introduction \label{sec:intro}}

To understand the formation of hierarchical structures in high-mass star formation regions, it is  critical to measure the dynamical coupling between density enhancements in giant molecular clouds and gas motion of their local environment
\citep{McKee2007-45, Motte2018-56, Henshaw2020-4}. Such studies may distinguish between competing concepts for high-mass star formation, such as monolithic collapse of turbulent cores in virial equilibrium \citep{McKee2003,Krumholz2007}, competitive accretion in a protocluster environment through Bondi-Hoyle accretion \citep{Bonnell1997,Bonnell2001}, turbulence-driven inertial-inflow \citep{Padoan2020} and gravity-dominated global hierarchical collapse \citep{Vazquez2009,Ballesteros2011,Hartmann2012,Vazquez2017,Vazquez2019}, as discussed in \citet{Zhou2022-514}. High-resolution observations show that density enhancements are organized in filamentary gas networks, especially in hub-filament systems. In such systems, converging flows are funneling matter into the hub through the filaments. 
Many case studies have suggested that hub-filament systems are the birth cradles of high-mass stars and clusters
\citep{Peretto2013,Henshaw2014,Zhang2015,Liu2016,Yuan2018,Lu2018,Issac2019,Dewangan2020,Liu2022-511,Zhou2022-514}. 
Numerical simulations of colliding flows and collapsing clumps reveal velocity gradients along the dense filamentary streams converging toward the hubs \citep{Wang2010,Gomez2014-791,Smith2016,Padoan2020}. In observations, velocity gradients along filaments are often interpreted as evidence for gas inflow along filaments \citep{Kirk2013,Liu2016,Yuan2018,Williams2018-613,Chen2019-875,Chen2020-891,Pillai2020-4,Zhou2022-514}. 

\citet{Zhou2022-514}
studied the physical properties and evolution of hub–filament systems in a large sample of protoclusters that were observed in the ATOMS (ALMA Three-millimeter Observations of Massive Star-forming regions) survey \citep{Liu2020}.  They found that hub-filament structures can exist not only in small–scale ($\sim$0.1 pc) dense cores but also in large–scale
clumps/clouds ($\sim$1-10 pc), suggesting that multi-scale hub-filament systems at various scales are common in regions forming massive stellar clusters in various Galactic environments.
The filaments in clumps observed in the ATOMS program show clear velocity gradients. The approximately symmetric distribution of positive and negative velocity gradients strongly indicates the existence of converging gas inflows along filaments. The observations confirm that high-mass stars in protoclusters may accumulate most of their mass through longitudinal inflow along filaments. 
Velocity and density fluctuations are discussed in detail in \citet{Henshaw2020-4}, who detected ubiquitous velocity fluctuations across all spatial scales and galactic environments and discovered oscillatory gas flows with wavelengths ranging from 0.3-400 pc that are coupled to regularly spaced density enhancements, probably formed via gravitational instabilities \citep{Henshaw2016-463, Elmegreen2018-863}. Furthermore, the
locations of some of these density enhancements spatially correlate with velocity gradient extrema, indicative of either convergent motion or collapse-induced rotation \citep{Clarke2016-458, Misugi2019-881}. 

In \citet{Zhou2022-514}, we measured the velocity gradients around the intensity peaks for all observed cores. The statistical analysis found that velocity gradients are very small at scales larger than $\sim$1 pc, probably suggesting the dominance of pressure–driven inertial inflow, which can originate either from large-scale turbulence or from cloud-scale gravitational contraction. Below $\sim$1 pc, velocity gradients dramatically increase as filament lengths decrease, indicating that the hub’s or core’s gravity dominates gas infall on small scales. 
Due to the FOV limitation of ALMA observation, our previous work was restricted to scales up to about 5 pc. In this paper, following a similar method as described in \citet{Zhou2022-514}, we can generalize the results from the clump-core scale to cloud-clump scale.

\begin{figure*}
\centering
\includegraphics[width=0.9\textwidth]{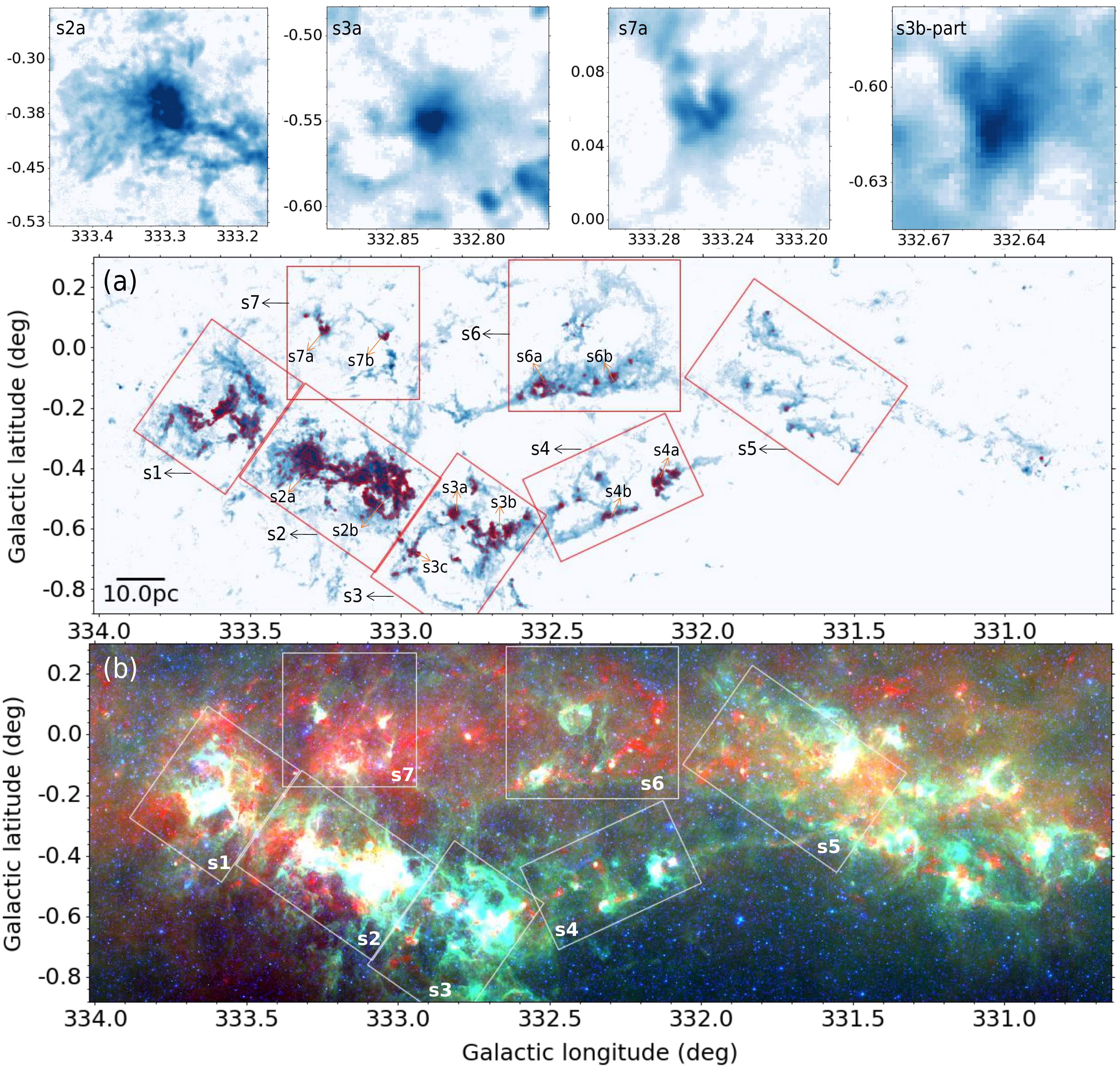}
\caption{Overview of the entire observed field.
The enlarged maps in the first row display several typical hub-filament structures. 
(a) Background is the integrated intensity map of $^{13}$CO (3-2) for the G333 complex, boxes show several divided sub-regions. Letters mark the sub-structures studied in this work. Red contours show the peak emission of $^{13}$CO (3-2); (b) An overview three-color map of the observed field by combining ATLASGAL+Planck 870 $\mu$m (red) and GLIMPSE 8.0 (green) and 4.5 $\mu$m (blue) emission, including G333 complex and G331 GMC.}
\label{sub}
\end{figure*}


Fig.\ref{sub}(b) displays an overview three-color map of the observed field by
combining ATLASGAL+Planck 870 $\mu$m and GLIMPSE 8.0 and 4.5 $\mu$m emission. Fig.\ref{sub}(a) shows the distribution of newly observed $^{13}$CO (3$-$2) emission. The interesting individual sub-regions are marked and highlighted below. The main structures of the observed field are the G331 and G333 giant molecular clouds (GMCs), and the G332 ring structure between them.

The G331 GMC is one of the most massive  molecular clouds in the Southern Galaxy, in the tangent region of the Norma spiral arm \citep{Bronfman1989-71}. Using the C$^{18}$O (1--0) integrated emission, \citet{Merello2013-774} defined the central region of the G331 GMC at $l=331.523\degr$, $b=-0.099\degr$, with a distance of 7.5\,kpc. 
It may harbor one of the most extended and luminous regions of massive star formation in the Galactic disk. 
\citet{Caswell1987-171} determined that the line central velocity of the ionized gas is $\sim-89$\,km\,s$^{-1}$ from observations of the H$109\alpha$ and H$110\alpha$ hydrogen recombination lines, which is similar to the peak velocity of the molecular gas in the GMC. The detection of OH and methanol maser emission further provides evidence of active star formation in the G331 GMC \citep{Goss1970-23, Caswell1980-33,Caswell1998-297,Pestalozzi2005-432}.

The G333 GMC, centered at $l \sim$ 333.2$\degr$, $b \sim-0.4\degr$, is a 1.2$\degr \times$ 0.6$\degr$ region in the fourth quadrant of the Galaxy at a distance of 3.6 kpc \citep{Lockman1979-232,Bains2006-367}. Its gas emission takes the form of a string of knots. 
As a part of the Galactic Ring of molecular clouds at a Galactocentric radius of 3–5 kpc, the G333 molecular cloud complex contains a diverse sample of molecular regions as well as a range of high mass star forming clouds, bright HII regions and infrared point sources, all surrounded by diffuse atomic and molecular gas \citep{Fujiyoshi2006-368,Cunningham2008-31,Wong2008-386,Lo2009-395,Jordan2013-429,Lowe2014-441}.
OH, H$_{2}$O and CH$_{3}$OH maser lines as tracers of high-mass star formation have been detected from various sources in the region \citep{Caswell1980-33, Caswell1995-277, Breen2007-377,Caswell2011-417,Breen2012-421}. 
Towards the G333 GMC, \citet{Bains2006-367} used the Mopra 22-m radio telescope to identify five distinct velocity features at $-105$, $-90$, $-70$, $-50$, and $-10$\,km\,s$^{-1}$ with $^{13}$CO (1$-$0) maps. They also found at least three velocity components ($-55$, $-50$, and $-42$\,km\,s$^{-1}$)  between $-65$ and $-35$\,km\,s$^{-1}$, with the brightest feature at $-50$\,km\,s$^{-1}$.  
Several spiral arms intersect at $l \sim$ 333$\degr$ along the line-of-sight, i.e., the Sagittarius-Carina arm, the Scutum-Crux arm and the Norma-Cygnus arm  \citep{Russeil2003-397,Russeil2005-429,Vallee2008-135}. This region may therefore be suitable to study the formation of molecular clouds under galaxy dynamics.

Between the G333 and G331 GMCs, there is a cavity at $l \sim$ 332$\degr$. The upper part presents a ring structure within Galactic coordinates  $332.0\degr< l < 332.8\degr$ and $-0.3\degr < b < 0.4\degr$. The principal emission of this ring structure is concentrated between the velocity range $-55\textless  v_{\rm LSR} \textless -44$\,km\,s$^{-1}$ \citep{Romano2019-484} and the average spectral profiles of the CO lines and CI show a peak at $v_{\rm LSR} = -50$\,km\,s$^{-1}$. The distance of the ring is determined to be $\sim$3.7\,kpc from the Sun \citep{Romano2019-484}, thus the ring and G333 GMC are located in the same region, referred to as the G333 complex in this work. 

\section{Observation}

\subsection{APEX/LAsMA data}

We mapped a $3.4\degr \times 1.2\degr$ area centered on $(l,b)=(332.33\degr,-0.29\degr)$. The observations were conducted between March and August of 2022 using the APEX telescope \citep{Gusten2006-454}\footnote{\tiny This publication is based on data acquired with the Atacama Pathfinder EXperiment (APEX). APEX is a collaboration between the Max-Planck-Institut f\"{u}r Radioastronomie, the European Southern Observatory and the Onsala Space Observatory.}. 
The 7 pixel Large APEX sub-Millimeter Array (LAsMA) receiver was used to observe the $J=3-2$ transitions of $^{12}$CO ($\nu_{\text{rest}}\sim345.796\,\si{\GHz}$) and $^{13}$CO ($\nu_{\text{rest}}\sim 330.588\,\si{\GHz}$) simultaneously in the upper and lower sideband, respectively. More details about the receiver are given in \citet{Mazumdar2021-650}. 
The local oscillator frequency was set at $\SI{338.190}{\GHz}$ in order to avoid contamination of the $^{13}$CO (3$-$2) lines due to bright $^{12}$CO (3$-$2) emission from the image band.
The whole observed region was divided into two parts, the first part is from $l \sim$ 332$\degr$ to $l \sim$ 334$\degr$, the second part is from $l \sim$ 330.6$\degr$ to $l \sim$ 332$\degr$. Each part was further divided into sub-maps of size $\SI{10}{\arcmin} \times \SI{10}{\arcmin}$ each. Observations were performed in a position switching on-the-fly (OTF) mode. Although the reference-positions were carefully chosen, the reference-position of the first part still has emission present in the velocity range of $-$140 to 10 km\,s$^{-1}$ (see Appendix\,\ref{app-a} for details about how the issue was solved). 

The data were calibrated using a three load chopper wheel method, which is an extension of the ``standard'' method used for millimeter observations \citep{Ulich1976-30} to calibrate the data to the antenna temperature $T^*_A$ scale. 
The data was reduced using the GILDAS package\footnote{\url{http://www.iram.fr/IRAMFR/GILDAS}}. A velocity range of $-$190 to 60 km\,s$^{-1}$ was extracted for each spectrum and resampled to an adequate velocity resolution of 0.25 km s$^{-1}$ to reduce the noise. The velocity range $-$140 to 10 km s$^{-1}$ was masked before fitting a first order baseline to each spectrum. The reduced, calibrated data obtained from the different scans were then combined and gridded using a $6 \arcsec$ cell size. The gridding process includes a convolution with a Gaussian kernel with a FWHM size of one-third the telescope FWHM beam width. The data cubes obtained have a final angular resolution of $19.5 \arcsec$ comparable with the resolution of the ATLASGAL survey. 

After the online calibration, the intensities are obtained on the $T^*_A$ (corrected antenna temperature) scale. Apart from the atmospheric attenuation, this also corrects for rear spillover, blockage, scattering and ohmic losses. A beam efficiency value $\eta_{mb}=0.71$ \citep{Mazumdar2021-650} was used to convert intensities from $T^*_A$ to the main beam brightness temperature $T_{mb}$. The final noise levels of the $^{12}$CO (3--2) and $^{13}$CO (3--2) data cubes are $\sim$0.32 K and $\sim$0.46 K, respectively. In the rest of paper, we focus on the analysis of the $^{13}$CO (3-2) emission. The information from both lines will be combined in a forthcoming paper.

\subsection{Archival continuum emission data }
The observed region was covered in the infrared by the GLIMPSE survey \citep{Benjamin2003}.
The images of the Spitzer Infrared Array Camera (IRAC) at 4.5 and 8.0 $\mu$m were retrieved from the Spitzer Archive. The angular resolutions of the images in the IRAC bands are $\sim 2\arcsec$.
We also used ATLASGAL+Planck 870 $\mu$m data (ATLASGAL combined with Planck data), which are sensitive to a wide range of spatial scales at a resolution of $\sim21\arcsec$ \citep{Csengeri2016-585}.

        \begin{table*}[h!]
        \caption{Properties of ATLASGAL clumps associated with the intensity peaks of $^{13}$CO (3$-$2) emission (orange circles in Fig.~\ref{example}), which are treated as hubs in this work. $\Delta$V and L represent the velocity gradient and the length of the filament. As shown in Fig.~\ref{example}, there are two velocity gradients around one intensity peak, marked as '1' and '2'. However, here we only reserve the good fittings, thus many hubs only have the velocity gradient of one side.}
        \label{clump}
        \centering
        \begin{tabular}{cccccccccc}
        \hline
        Clump	&	log(M)	&	Radius	&	T$_{\rm{dust}}$	&	log(Lum)	&	v$_{\rm{LSR}}$	&	$\Delta$V$_{\rm{1}}$	&	L$_{\rm{1}}$	&	$\Delta$V$_{\rm{2}}$	&	L$_{\rm{2}}$	\\
        &($\mathrm{M_{\odot}}$) & ($\mathrm{pc}$) & ($\mathrm{K}$) & ($\mathrm{L_{\odot}}$) & ($\mathrm{km\,s^{-1}}$) & ($\mathrm{km\,s^{-1}\,pc^{-1}}$) & ($\mathrm{pc}$) & ($\mathrm{km\,s^{-1}\,pc^{-1}}$) & ($\mathrm{pc}$) \\
        \hline \\
AGAL332.094$-$00.421	&	3.155	&	0.865	&	28.4	&	4.789	&	$-$56.5	&	$-$2.97	&	1.4 	&	1.9 	&	1.1 	\\
AGAL332.312$-$00.556	&	2.947	&	0.831	&	21.4	&	3.835	&	$-$51.9	&	0.9 	&	1.9 	&	$-$1.76	&	0.6 	\\
AGAL332.351$-$00.436	&	3.034	&	0.935	&	18.3	&	3.483	&	$-$47.6	&		&		&	$-$0.58	&	6.0 	\\
AGAL332.411$-$00.471	&	2.618	&	0.502	&	19.3	&	2.694	&	$-$55.8	&	0.3 	&	6.0 	&	$-$0.49	&	5.0 	\\
AGAL332.467$-$00.522	&	3.162	&	0.848	&	19.8	&	3.864	&	$-$52.6	&	0.2 	&	3.0 	&	$-$2.25	&	0.8 	\\
AGAL332.606$-$00.854	&	2.569	&	0.537	&	19.8	&	3.088	&	$-$56.2	&	$-$0.53	&	2.0 	&	2.0 	&	1.0 	\\
AGAL332.647$-$00.609	&	3.549	&	1.35	&	28.3	&	5.195	&	$-$49.2	&	$-$2.7	&	2.0 	&	3.7 	&	1.0 	\\
AGAL332.694$-$00.612	&	3.184	&	0.813	&	26.6	&	4.466	&	$-$47.6	&	$-$7.84	&	1.0 	&	9.7 	&	1.0 	\\
AGAL332.694$-$00.612	&	3.184	&	0.813	&	26.6	&	4.466	&	$-$47.6	&	3.3 	&	0.8 	&	$-$4.59	&	1.0 	\\
AGAL332.726$-$00.621	&	2.407	&	0.294	&	24.8	&	3.456	&	$-$49.7	&	$-$3.2	&	0.7 	&	1.5 	&	0.8 	\\
AGAL332.751$-$00.597	&	2.964	&	1.038	&	24.7	&	3.937	&	$-$53.1	&		&		&	0.8 	&	1.8 	\\
AGAL332.762$-$00.641	&	2.32	&	0.623	&	22.4	&	3.313	&	$-$50.4	&	$-$4.14	&	0.6 	&	0.9 	&	1.7 	\\
AGAL332.826$-$00.549	&	3.724	&	1.367	&	31.4	&	5.543	&	$-$57.3	&	$-$0.58	&	6.7 	&	2.7 	&	1.0 	\\
AGAL332.826$-$00.549	&	3.724	&	1.367	&	31.4	&	5.543	&	$-$57.3	&	$-$2.65	&	2.5 	&	3.1 	&	1.0 	\\
AGAL332.826$-$00.549	&	3.724	&	1.367	&	31.4	&	5.543	&	$-$57.3	&		&		&	$-$0.86	&	8.5 	\\
AGAL332.892$-$00.569	&	2.595	&	0.485	&	19.7	&	3.19	&	$-$57.3	&	$-$1.49	&	1.5 	&	0.9 	&	2.0 	\\
AGAL332.866$-$00.587	&	2.364	&	0.169	&	23.3	&	3.113	&	$-$58.1	&		&		&	$-$1.59	&	1.0 	\\
AGAL332.962$-$00.679	&	3.305	&	0.779	&	22.5	&	4.225	&	$-$48.5	&	2.0 	&	3.0 	&	$-$1.53	&	5.5 	\\
AGAL332.969$-$00.737	&	2.785	&	0.64	&	13.5	&	2.525	&	$-$55.6	&	$-$3.7	&	1.0 	&	3.9 	&	1.0 	\\
AGAL332.995$-$00.519	&	2.511	&	0.467	&	18.2	&	2.593	&	$-$52.8	&		&		&	3.3 	&	1.0 	\\
AGAL333.001$-$00.436	&	2.848	&	0.865	&	35.7	&	4.989	&	$-$55.4	&	$-$3.4	&	0.7 	&	1.0 	&	2.5 	\\
AGAL333.013$-$00.466	&	2.821	&	0.606	&	34.5	&	4.968	&	$-$53.3	&	$-$1.74	&	1.2 	&	3.4 	&	0.6 	\\
AGAL333.014$-$00.521	&	2.986	&	1.056	&	22.6	&	4.152	&	$-$53.2	&	$-$2.58	&	1.0 	&	1.3 	&	2.0 	\\
AGAL333.053+00.029	&	3.017	&	1.056	&	29.2	&	4.398	&	$-$45.3	&	1.1 	&	1.4 	&	$-$0.42	&	2.1 	\\
AGAL333.071$-$00.399	&	3.281	&	0.969	&	19.1	&	3.744	&	$-$53.3	&		&		&	0.9 	&	1.2 	\\
AGAL333.071$-$00.399	&	3.281	&	0.969	&	19.1	&	3.744	&	$-$53.3	&	$-$1.14	&	1.5 	&	0.2 	&	9.0 	\\
AGAL333.089$-$00.352	&	2.347	&	0.692	&	28.1	&	3.95	&	$-$52.7	&		&		&	0.6 	&	2.5 	\\
AGAL333.094$-$00.524	&	2.728	&	0.623	&	26.5	&	3.966	&	$-$57.6	&	$-$1.64	&	1.5 	&	0.8 	&	3.5 	\\
AGAL333.103+00.087	&	2.071	&	0.169	&	17	&	2.119	&	$-$44.9	&	0.3 	&	2.0 	&	$-$0.26	&	3.0 	\\
AGAL333.103$-$00.502	&	3.251	&	1.367	&	25.5	&	4.601	&	$-$55.8	&	$-$0.81	&	6.0 	&	1.3 	&	1.0 	\\
AGAL333.129$-$00.559	&	3.843	&	1.402	&	16.1	&	3.983	&	$-$56.4	&	$-$0.8	&	4.0 	&	2.4 	&	2.0 	\\
AGAL333.134$-$00.431	&	3.959	&	1.506	&	32	&	5.823	&	$-$51.9	&	$-$3.3	&	1.3 	&	2.0 	&	1.7 	\\
AGAL333.169$-$00.431	&	2.959	&	0.831	&	23	&	4.052	&	$-$50.8	&	$-$0.63	&	2.2 	&	2.0 	&	1.3 	\\
AGAL333.248+00.054	&	3.27	&	1.454	&	26.3	&	4.706	&	$-$46.9	&		&		&	$-$0.81	&	5.5 	\\
AGAL333.264$-$00.291	&	2.229	&	0.485	&	27.4	&	3.951	&	$-$52.2	&		&		&	$-$3.4	&	0.8 	\\
AGAL333.284$-$00.387	&	3.678	&	1.211	&	32.1	&	5.435	&	$-$51.6	&		&		&	0.7 	&	3.0 	\\
AGAL333.284$-$00.387	&	3.678	&	1.211	&	32.1	&	5.435	&	$-$51.6	&	$-$1.66	&	1.8 	&	2.5 	&	1.0 	\\
AGAL333.308$-$00.366	&	3.561	&	1.246	&	34	&	5.544	&	$-$50.2	&	$-$4.51	&	0.4 	&	0.9 	&	3.8 	\\
AGAL333.308$-$00.366	&	3.561	&	1.246	&	34	&	5.544	&	$-$50.2	&		&		&	2.8 	&	1.0 	\\
AGAL333.308$-$00.366	&	3.561	&	1.246	&	34	&	5.544	&	$-$50.2	&	$-$2.5	&	1.2 	&	2.8 	&	0.7 	\\
AGAL333.308$-$00.366	&	3.561	&	1.246	&	34	&	5.544	&	$-$50.2	&	$-$0.32	&	6.0 	&	0.2 	&	6.0 	\\
AGAL333.314+00.106	&	2.901	&	0.71	&	23.3	&	3.922	&	$-$47.7	&	2.4 	&	1.3 	&	$-$0.88	&	2.7 	\\
AGAL333.466$-$00.164	&	3.427	&	1.038	&	24.6	&	4.591	&	$-$42.5	&	2.3 	&	1.5 	&	$-$1.57	&	5.0 	\\
AGAL333.483$-$00.246	&	3.133	&	0.623	&	18	&	3.255	&	$-$50.4	&		&		&	$-$1.57	&	5.0 	\\
AGAL333.499$-$00.329	&	2.163	&	0.433	&	23.9	&	3.244	&	$-$50.2	&		&		&	$-$2.18	&	1.0 	\\
AGAL333.506$-$00.101	&	2.653	&	0.744	&	19.7	&	3.096	&	$-$42.8	&		&		&	1.3 	&	2.0 	\\
AGAL333.524$-$00.269	&	3.488	&	1.54	&	22.5	&	4.252	&	$-$48.8	&	$-$0.42	&	3.5 	&	4.9 	&	1.3 	\\
AGAL333.524$-$00.269	&	3.488	&	1.54	&	22.5	&	4.252	&	$-$48.8	&	$-$1.41	&	0.7 	&		&		\\
AGAL333.539$-$00.242	&	2.93	&	0.727	&	22.8	&	3.873	&	$-$48.9	&	$-$0.79	&	2.5 	&	1.5 	&	1.0 	\\
AGAL333.566$-$00.296	&	2.982	&	1.09	&	21.7	&	3.881	&	$-$46.7	&	$-$2.56	&	1.9 	&	1.3 	&	3.3 	\\
AGAL333.651$-$00.334	&	2.832	&	0.796	&	18.2	&	3.083	&	$-$47.1	&	$-$1.51	&	3.5 	&	1.6 	&	1.0 	\\
AGAL333.683$-$00.256	&	2.83	&	0.883	&	28.3	&	4.18	&	$-$48.1	&	$-$2.83	&	1.0 	&	0.9 	&	1.5 	\\
AGAL333.696$-$00.231	&	2.519	&	0.398	&	29.7	&	3.653	&	$-$46.7	&		&		&	$-$0.29	&	4.8 	\\
AGAL333.774$-$00.256	&	2.993	&	0.692	&	15.5	&	3.061	&	$-$49.5	&	$-$2.7	&	1.3 	&	1.0 	&	0.7 	\\
AGAL331.706$-$00.266	&	2.674	&	0.661	&	16.1	&	2.631	&	$-$49	&		&		&	$-$0.92	&	1.6 	\\
AGAL331.856$-$00.126	&	3.05	&	0.947	&	18.2	&	3.539	&	$-$50.1	&		&		&	$-$0.39	&	3.2 	\\
        \hline
        \end{tabular}
        \end{table*}     

        \begin{table*}[h!]
        \caption{Remaining part of Table\,\ref{clump}. 
        Usually one clump corresponds to one hub.
        Here 2 or 3 clumps separated by the gap indicate that the 2 or 3 clumps together form a hub.}
        \label{clump1}
        \centering
        \begin{tabular}{cccccccccc}
        \hline
        Clump	&	log(Mass)	&	Radius	&	T$_{\rm{dust}}$	&	log(Lum)	&	v$_{\rm{LSR}}$	&	$\Delta$v$_{\rm{1}}$	&	L$_{\rm{1}}$	&	$\Delta$v$_{\rm{2}}$	&	L$_{\rm{2}}$	\\
        &($\mathrm{M_{\odot}}$) & ($\mathrm{pc}$) & ($\mathrm{K}$) & ($\mathrm{L_{\odot}}$) & ($\mathrm{km\,s^{-1}}$) & ($\mathrm{km\,s^{-1}\,pc^{-1}}$) & ($\mathrm{pc}$) & ($\mathrm{km\,s^{-1}\,pc^{-1}}$) & ($\mathrm{pc}$) \\
        \hline \\
AGAL332.166+00.126	&	2.515	&	0.541	&	14.4	&	2.316	&	$-$49.9	&	$-$0.65	&	2.0 	&	0.7 	&	3.5 	\\
AGAL332.226$-$00.124	&	2.892	&	0.556	&	15.4	&	2.579	&	$-$50.8	&	$-$2.63	&	1.0 	&	1.6 	&	2.0 	\\
AGAL332.226$-$00.124	&	2.892	&	0.556	&	15.4	&	2.579	&	$-$50.8	&	$-$1.57	&	1.5 	&	2.9 	&	0.7 	\\
AGAL332.241$-$00.044	&	3.226	&	0.917	&	17	&	3.526	&	$-$48.1	&	$-$1.45	&	1.0 	&	3.0 	&	1.0 	\\
AGAL332.254$-$00.056	&	2.862	&	0.421	&	14.3	&	2.675	&	$-$47.7	&		&		&	0.8 	&	1.7 	\\
AGAL332.296$-$00.094	&	3.15	&	0.932	&	24	&	4.303	&	$-$48.2	&	0.6 	&	3.0 	&	$-$0.36	&	2.5 	\\
AGAL332.317+00.177	&	2.576	&	0.406	&	13.6	&	2.223	&	$-$48.6	&		&		&	2.0 	&	0.5 	\\
AGAL332.352$-$00.116	&	2.513	&	0.601	&	27.9	&	3.887	&	$-$49.8	&	$-$0.39	&	5.0 	&	0.5 	&	2.0 	\\
AGAL332.442$-$00.139	&	2.723	&	0.421	&	18.3	&	2.669	&	$-$51.3	&		&		&	0.7 	&	2.0 	\\
AGAL332.469$-$00.131	&	2.643	&	0.556	&	20.3	&	3.088	&	$-$50.5	&	1.8 	&	1.2 	&	$-$1.3	&	1.2 	\\
AGAL332.544$-$00.124	&	2.857	&	1.037	&	32.5	&	4.591	&	$-$47.5	&		&		&	$-$1.92	&	2.4 	\\
AGAL332.604$-$00.167	&	3.185	&	1.052	&	17.7	&	3.371	&	$-$46.2	&		&		&	$-$0.89	&	1.6 	\\
	&		&		&		&		&		&		&		&		&		\\
AGAL332.141$-$00.466	&	2.66	&	0.571	&	24.2	&	3.495	&	$-$57	&		&		&		&		\\
AGAL332.144$-$00.469	&	2.711	&	0.883	&	24.5	&	4.203	&	$-$58	&		&		&	3.0 	&	0.8 	\\
	&		&		&		&		&		&		&		&		&		\\
AGAL332.141$-$00.446	&	2.883	&	0.71	&	27.7	&	4.76	&	$-$56.3	&		&		&		&		\\
AGAL332.147$-$00.439	&	2.371	&	0.169	&	26.9	&	3.974	&	$-$57.9	&		&		&		&		\\
AGAL332.156$-$00.449	&	3.239	&	1.004	&	30	&	5.078	&	$-$55.6	&		&		&	3.0 	&	0.9 	\\
	&		&		&		&		&		&		&		&		&		\\
AGAL332.252$-$00.539	&	2.821	&	0.45	&	17.8	&	2.608	&	$-$52.9	&		&		&		&		\\
AGAL332.281$-$00.547	&	3.133	&	0.71	&	16.9	&	3.299	&	$-$52.2	&	$-$1.73	&	1.0 	&	2.2 	&	1.0 	\\
	&		&		&		&		&		&		&		&		&		\\
AGAL333.521$-$00.241	&	2.713	&	0.415	&	20	&	2.93	&	$-$48.8	&		&		&		&		\\
AGAL333.524$-$00.269	&	3.488	&	1.54	&	22.5	&	4.252	&	$-$48.8	&	2.3 	&	1.0 	&	$-$1.36	&	1.0 	\\
	&		&		&		&		&		&		&		&		&		\\
AGAL332.442$-$00.139	&	2.723	&	0.421	&	18.3	&	2.669	&	$-$51.3	&		&		&		&		\\
AGAL332.469$-$00.131	&	2.643	&	0.556	&	20.3	&	3.088	&	$-$50.5	&	$-$0.74	&	2.0 	&	0.7 	&	4.5 	\\
	&		&		&		&		&		&		&		&		&		\\
AGAL332.751$-$00.597	&	2.964	&	1.038	&	24.7	&	3.937	&	$-$53.1	&		&		&		&		\\
AGAL332.774$-$00.584	&	3.103	&	1.038	&	23	&	4.183	&	$-$55.2	&	$-$1.74	&	3.3 	&	5.6 	&	1.0 	\\
        \hline
        \end{tabular}
        \end{table*}  

\section{Results}
\subsection{Velocity components}\label{components}

\begin{figure}
\centering
\includegraphics[width=0.47\textwidth]{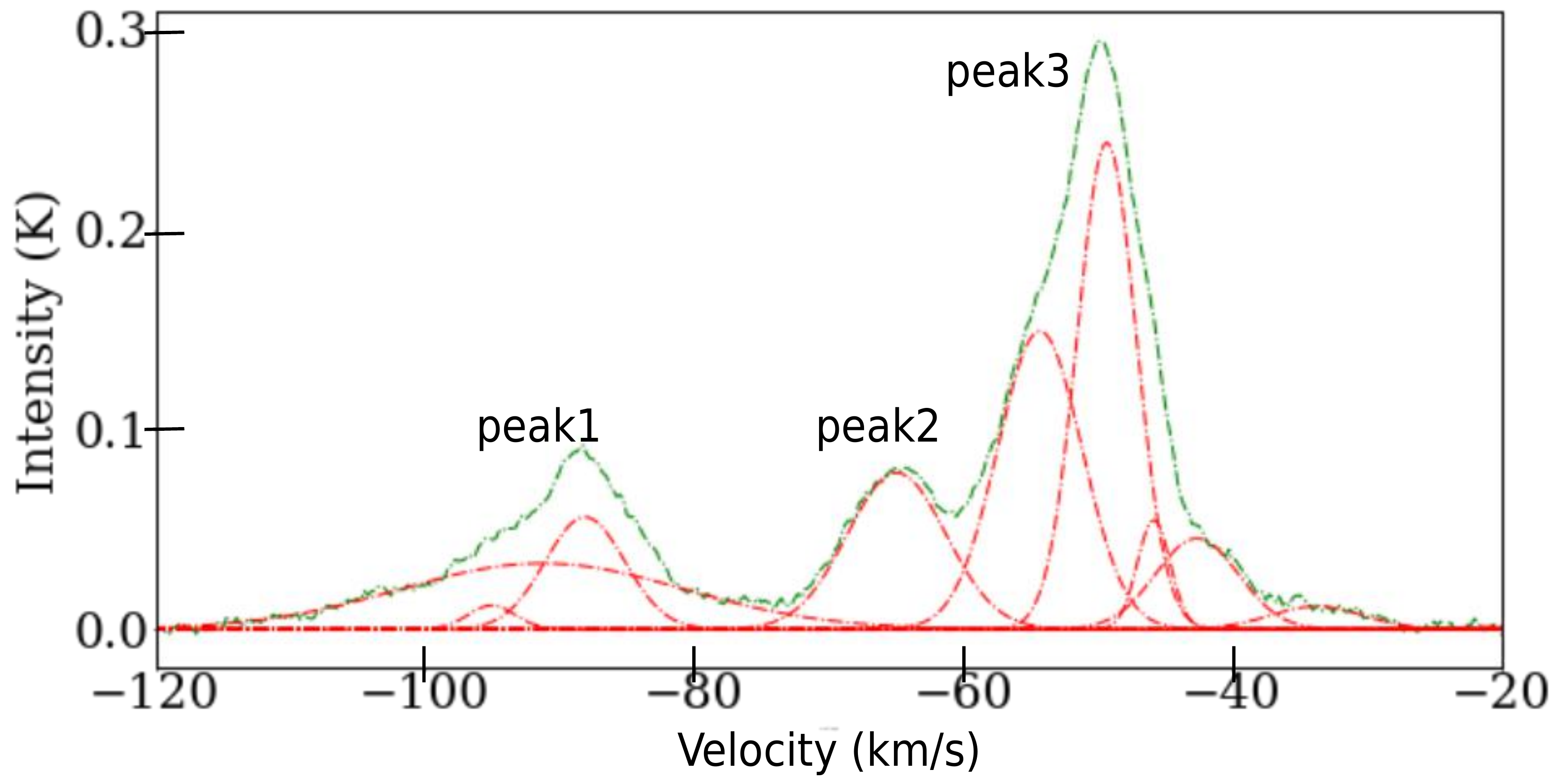}
\caption{
Average spectra of $^{13}$CO (3$-$2) for the entire region, red dashed profiles represent the components of multi-Gaussian fitting.
}
\label{spectra}
\end{figure}

\begin{figure*}
\centering
\includegraphics[width=0.8\textwidth]{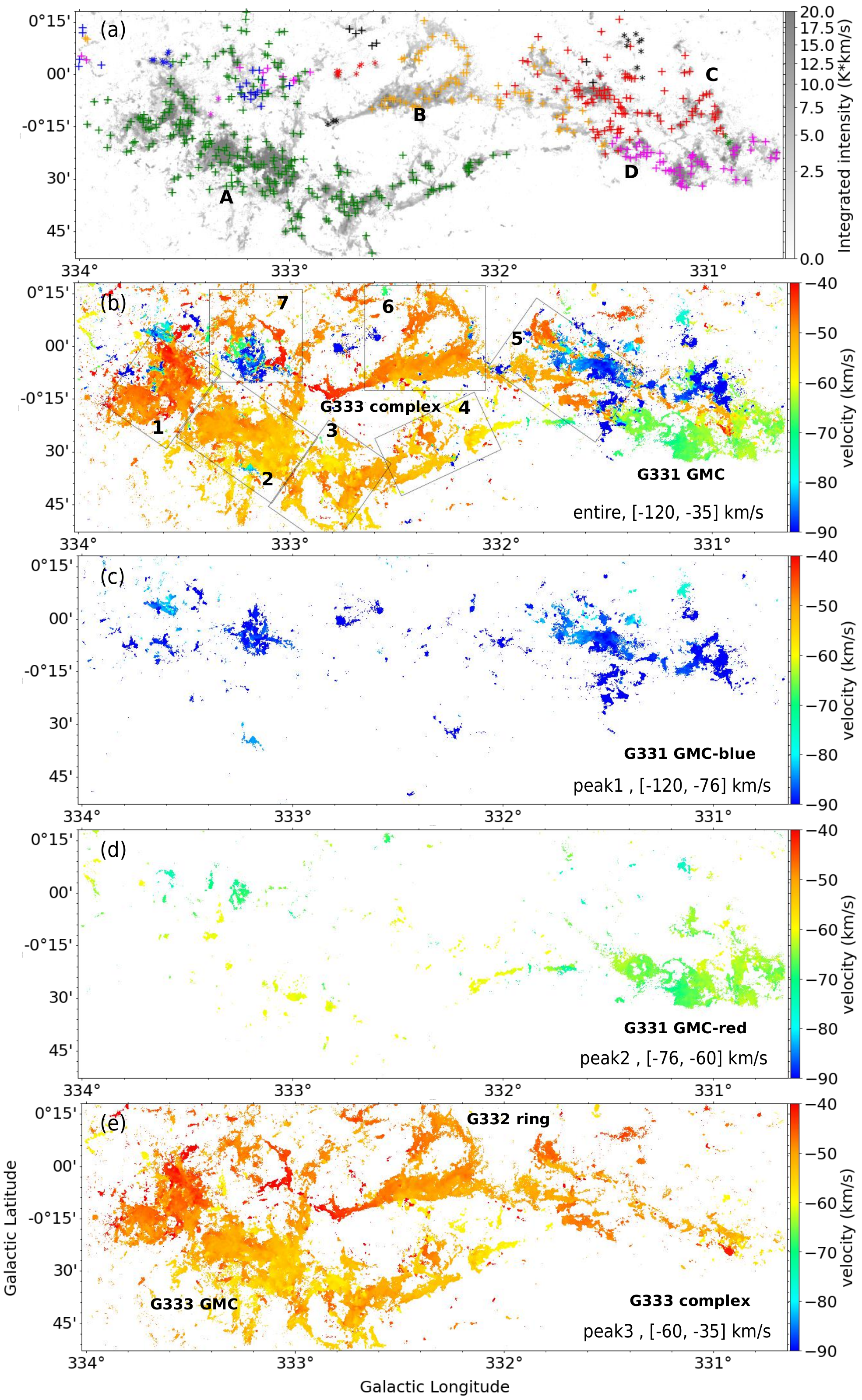}
\caption{ATLASGAL clumps distribution and velocity components of the entire observed field.
(a) Background is the integrated intensity map of $^{13}$CO (3$-$2), color-coded "+" mark different ATLASGAL clump clusters A (green), B (orange), C (red) and D (magenta); (b) The moment-1 map of $^{13}$CO (3$-$2) for the entire region in the overall velocity range; (c), (d) and (e) The moment-1 map of $^{13}$CO (3$-$2) for three peaks marked in Fig.\,\ref{spectra}.}
\label{large}
\end{figure*}

\begin{figure*}
\centering
\includegraphics[width=0.9\textwidth]{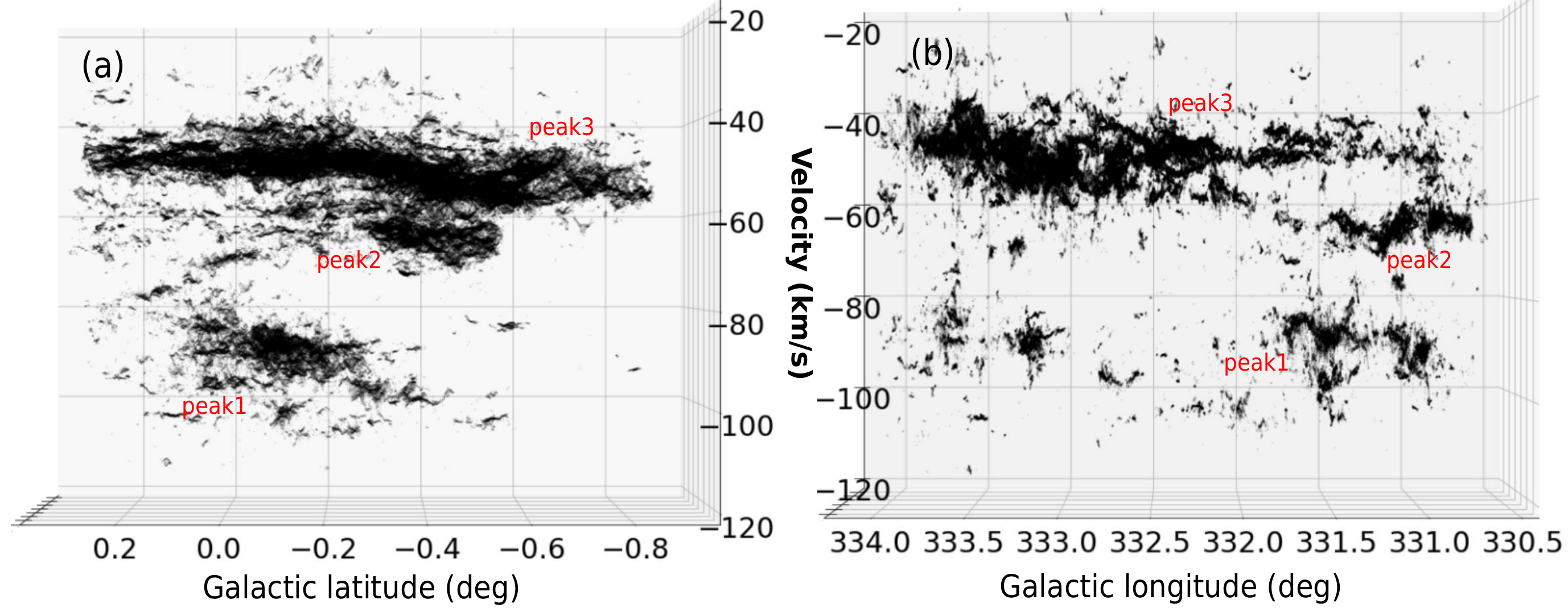}
\caption{The velocity distribution of the entire observed field in PPV space decomposed by \texttt{GAUSSPY+} along the Galactic latitude and longitude.}
\label{ppv}
\end{figure*}

\begin{figure*}
\centering
\includegraphics[width=0.9\textwidth]{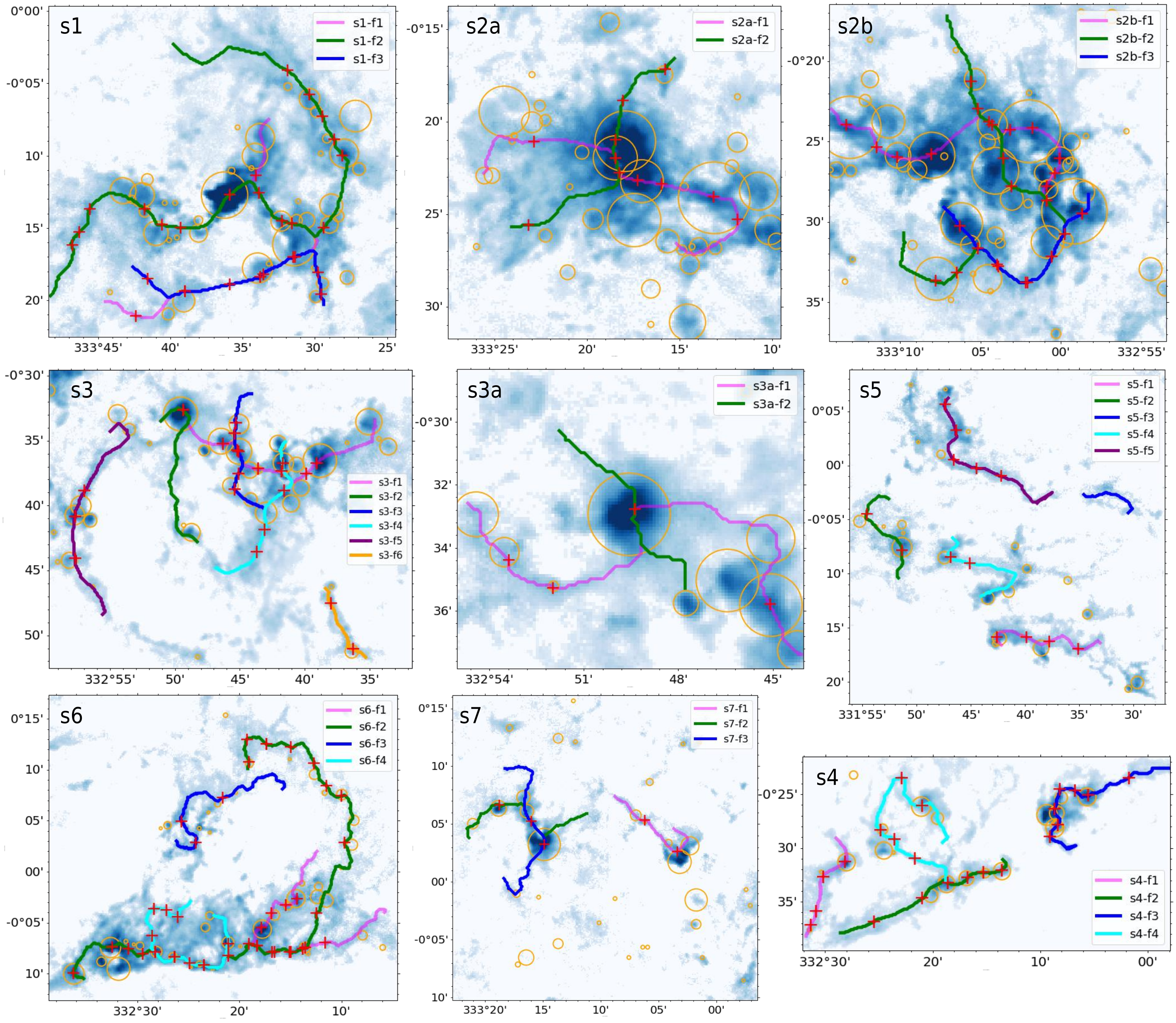}
\caption{Background images show the moment 0 maps of $^{13}$CO (3-2). Lines in color present the filament skeletons. Orange circles show the ATLASGAL clumps, the size of the circle represents the clump radius. Red '+' marks the intensity peak of $^{13}$CO (3-2) emission in Fig.~\ref{example}.}
\label{fil}
\end{figure*}

\begin{figure*}
\centering
\includegraphics[width=0.8\textwidth]{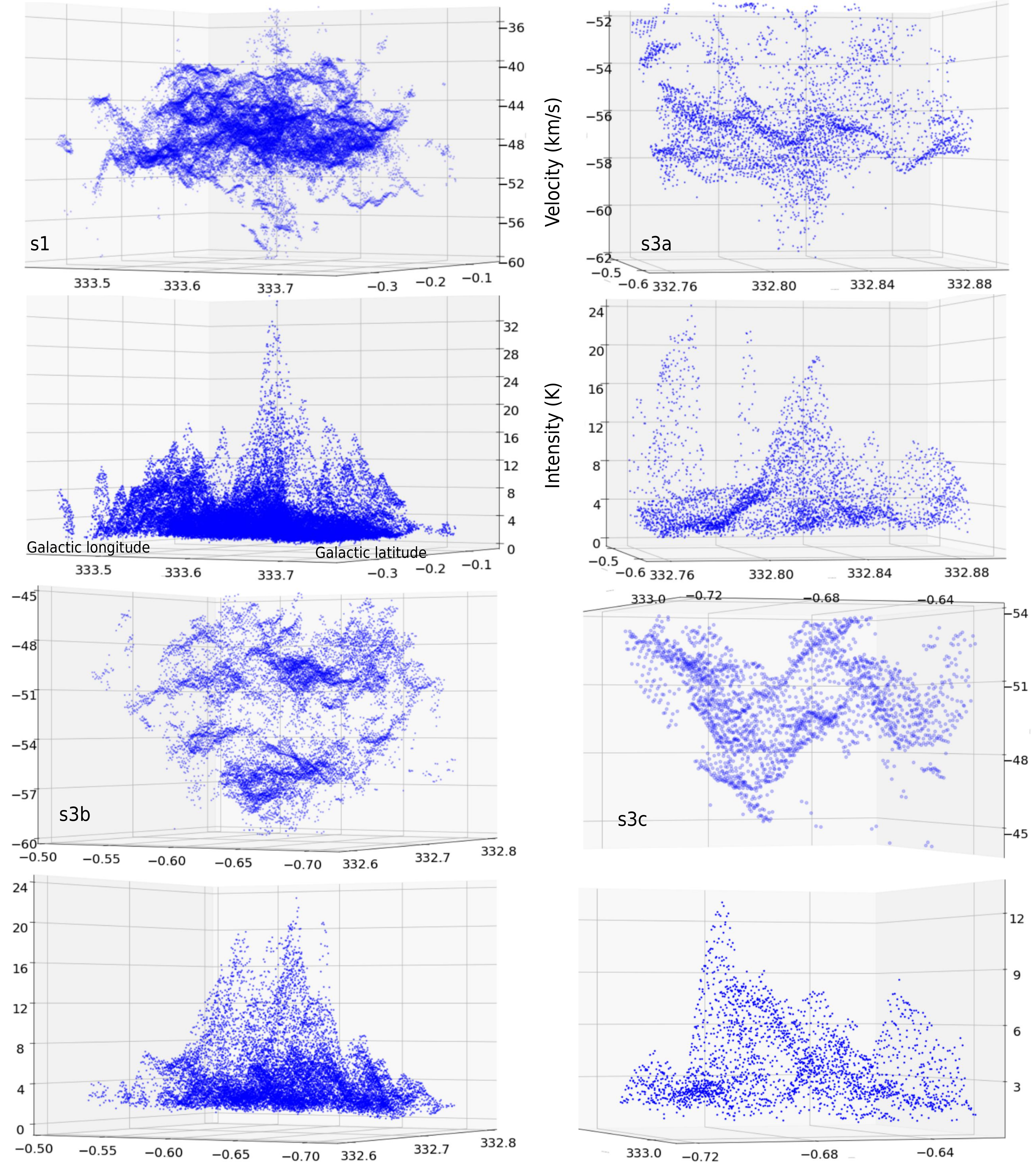}
\caption{The intensity and velocity distribution of sub-regions in PPV space based on the decomposition of \texttt{GAUSSPY+}. For other sub-regions, see Fig.~\ref{hub2} and Fig.~\ref{hub3} in
Sec.~\ref{supply}.}
\label{hub}
\end{figure*}

\begin{figure*}
\centering
\includegraphics[width=0.9\textwidth]{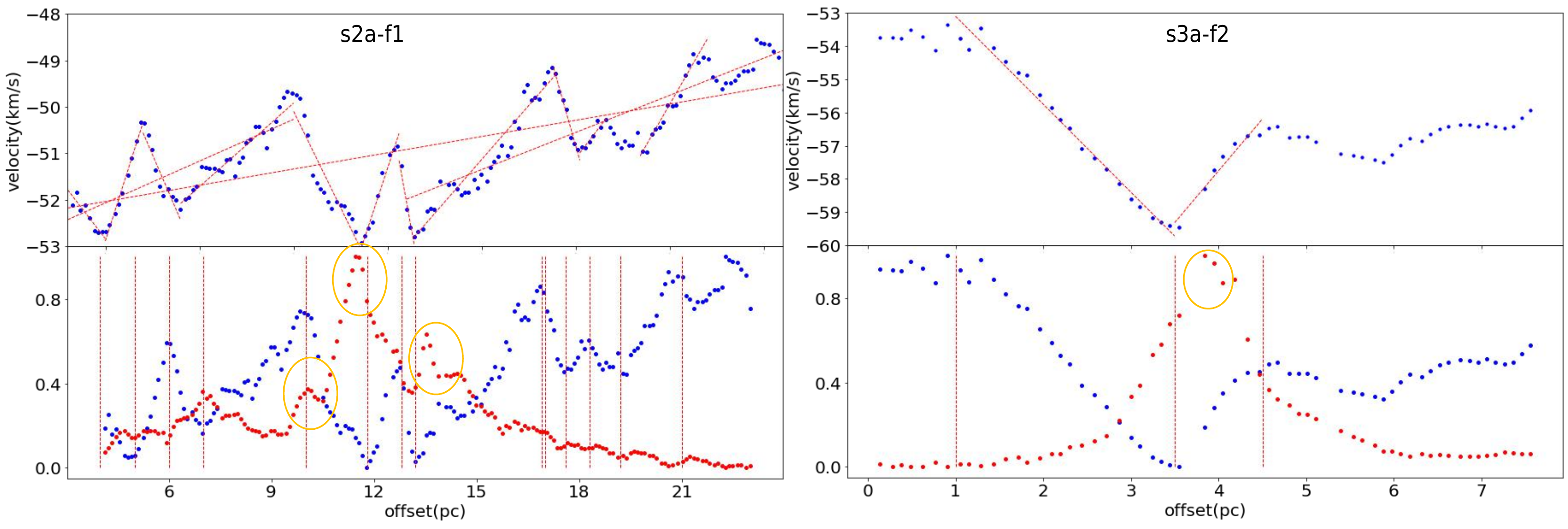}
\caption{Two selected filaments are used to illustrate the fitting of the velocity gradient.  Upper panel: Velocity gradients are fitted in the ranges defined by the red vertical dashed lines, and straight lines show the linear fitting results. Lower panel: Blue and red dotted lines show the normalized velocity and intensity, respectively. Orange circles mark the intensity peaks of $^{13}$CO (3$-$2) emission associated with ATLASGAL clumps.
For other filaments, see Fig.\,\ref{example1}, Fig.\,\ref{example2}, Fig.\,\ref{example3}, Fig.\,\ref{example4} and Fig.\,\ref{example5} in Sec.\,\ref{supply}.}
\label{example}
\end{figure*}

\begin{figure}
\centering
\includegraphics[width=0.47\textwidth]{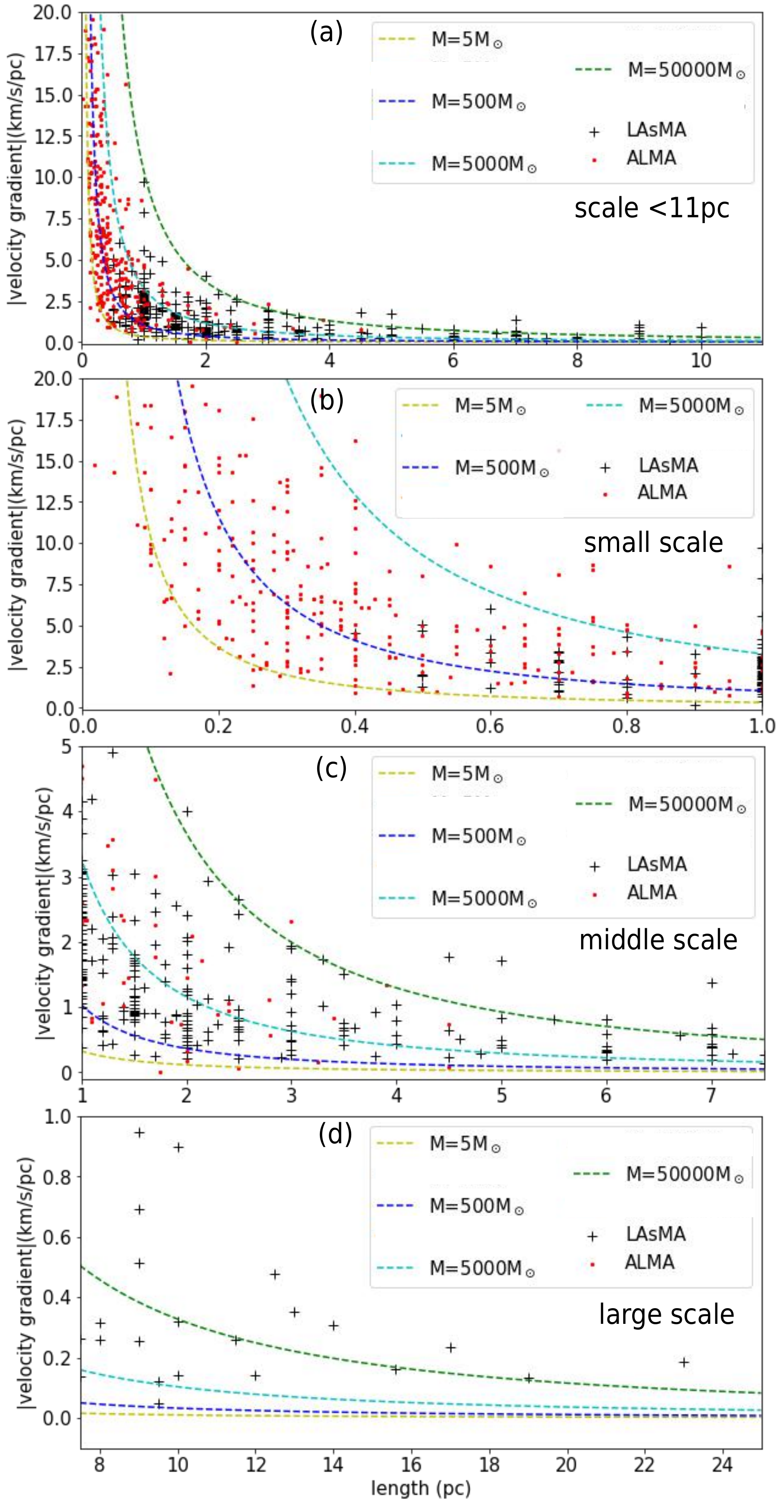}
\caption{ Statistical analysis of all fitted velocity gradients.
(a) Velocity gradient versus the length over which the gradient has been estimated for all sources in ATOMS survey (red dots) and current LAsMA observation (black '+'). The color lines show free-fall velocity gradients for comparison. For the free-fall model, yellow, blue, cyan and green lines denote masses of 5\,M$_\odot$, 500\,M$_\odot$, 5000\,M$_\odot$ and 50000M$_\odot$, respectively;
(b), (c) and (d) Blow-up maps with lengths $\textless$ 1\,pc (small scale), $\sim$ 1 -- 7.5\,pc (middle scale) and $\textgreater$ 7.5\,pc (large scale) in panel (a). 
}
\label{gradient}
\end{figure}

\begin{figure*}
\centering
\includegraphics[width=0.8\textwidth]{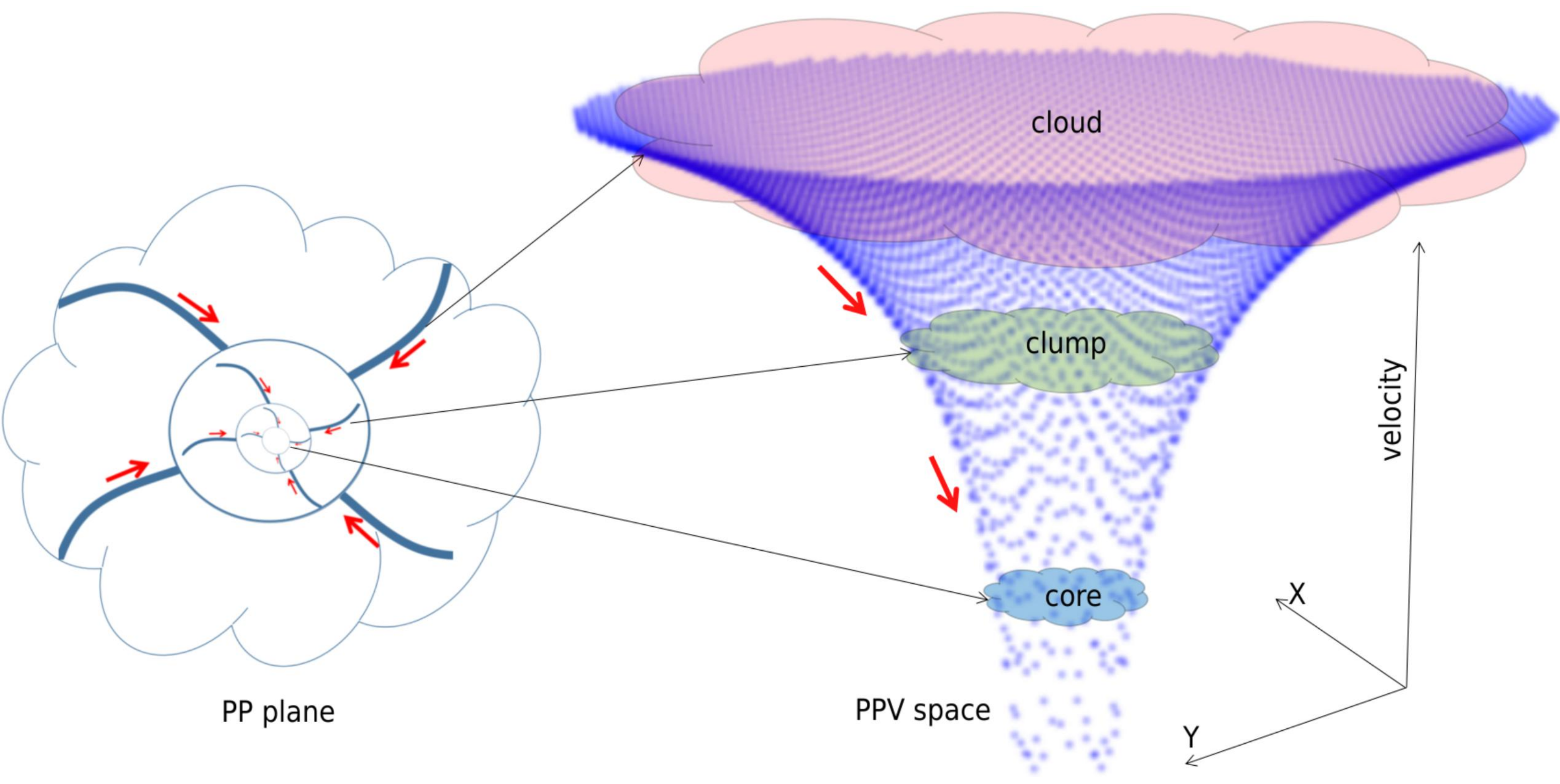}
\caption{Schematic diagram of the `funnel' structure in PPV space and on PP plane. PP plane shows a multi-scale hub-filament structure in a molecular cloud, which can turn to a ‘funnel’ structure in PPV space.
Red arrow represents the gas inflow.}
\label{model}
\end{figure*}

\begin{figure}
\centering
\includegraphics[width=0.47\textwidth]{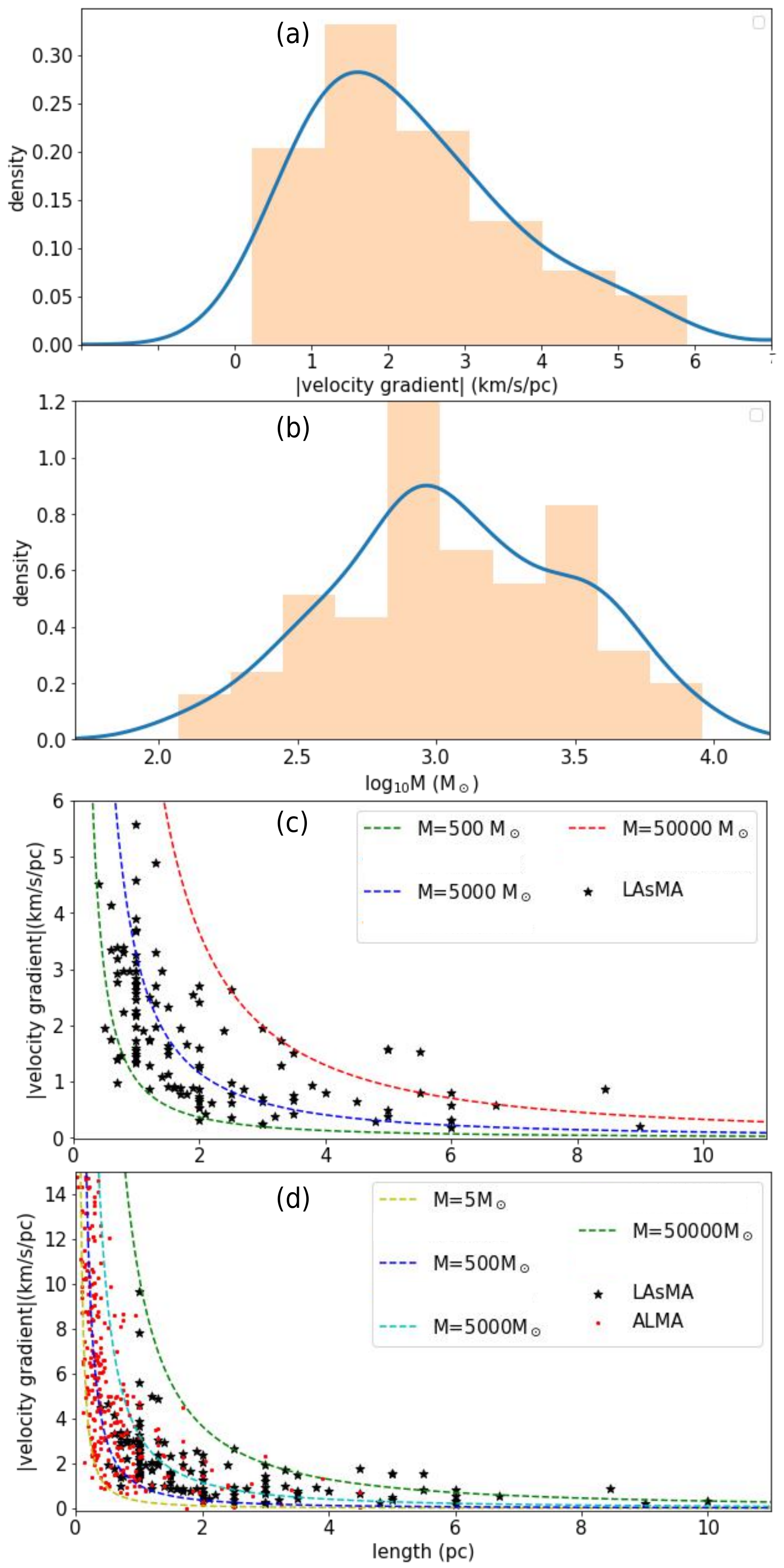}
\caption{
Statistical analysis of the velocity gradients listed in Table\,\ref{clump}.
(a) The probability distribution of velocity gradients measured around 1\,pc scale; (b) Mass distribution of ATLASGAL clumps listed in Table\,\ref{clump}; (c) and (d) The same with Fig.\,\ref{gradient}, but only consider the velocity gradients listed in Table\,\ref{clump}.}
\label{mas}
\end{figure}

\begin{figure}
\centering
\includegraphics[width=0.47\textwidth]{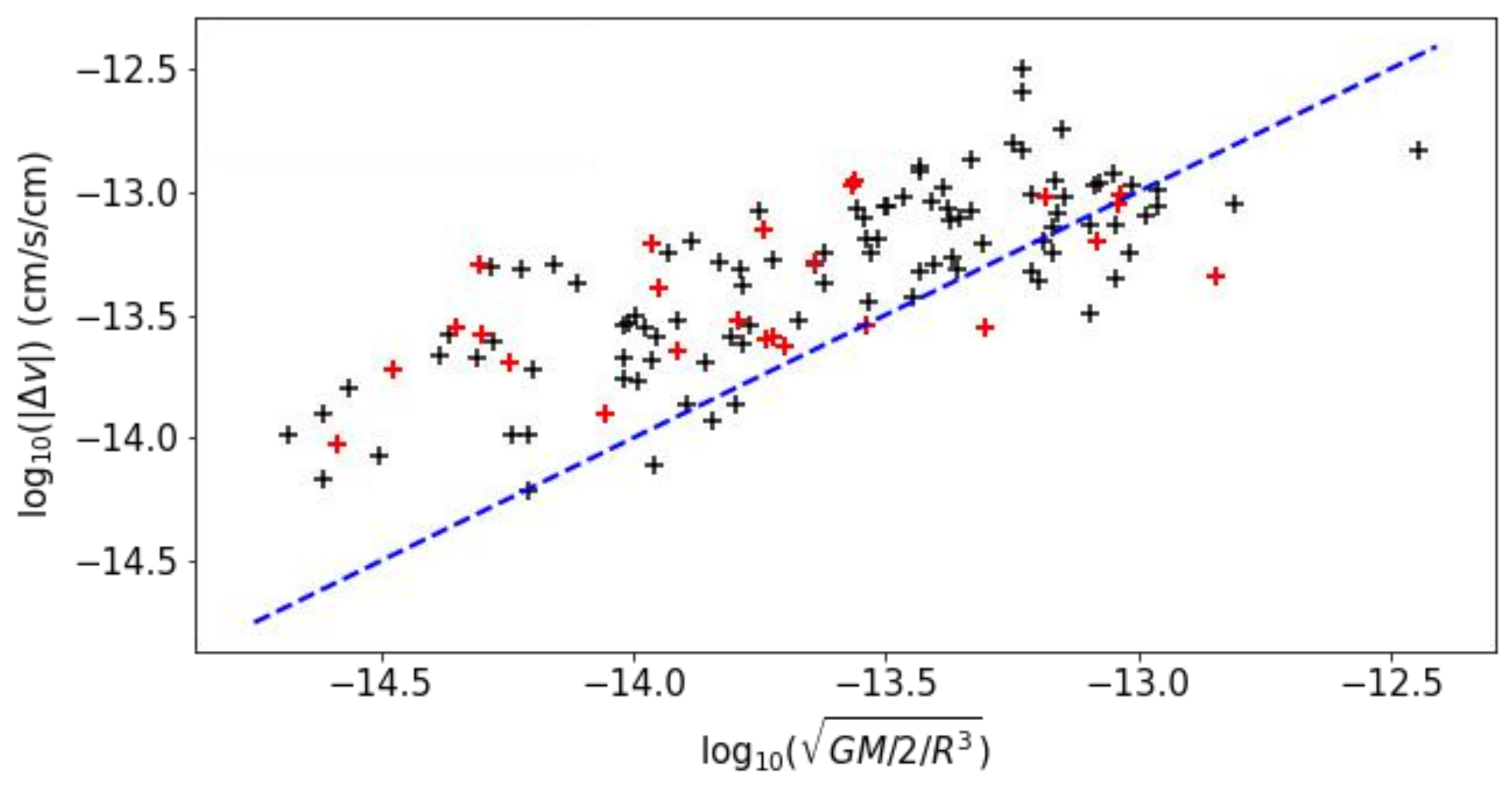}
\caption{Correlation between the left and right terms in Eqn.\,\ref{v-free} revealed by the values of $M$, $L$ and $\nabla V$ listed in Table\,\ref{clump}. Blue dotted line is the line of equality
of the left and right terms. There are two cases in Table\,\ref{clump}, a dense clump has two or more filaments leading to it, or only one filament with a gradient can be identified (the rows with only one entry for the gradient). The number of clumps in two cases is 49 and 25, and marked by black and red '+', respectively.}
\label{free}
\end{figure}

The average emission spectrum of $^{13}$CO (3$-$2) coming from the entire observed field is shown in Fig.\,\ref{spectra}, together with Gaussian curves used to fit the average spectrum. The fitting is carried out using the \texttt{GAUSSPY+} software \citep{Riener2019-628}, which will be introduced in detail below. We have divided the spectrum into three velocity components [$-$120, $-$76] km s$^{-1}$, [$-$76, $-$60] km s$^{-1}$ and [$-$60, $-$35] km s$^{-1}$ in which emission is prominent, marked as peak1, peak2 and peak3. 
From Fig.\,\ref{large}, we can see the velocity components peak1 and peak2 represent the main two parts of the G331 GMC, called G331 GMC-blue and G331 GMC-red.
peak3 shows a very extended structure all over the observed region with the G333 GMC and G332 ring as the main parts. 
There are four main ATLASGAL clump clusters marked by A, B, C and D in Fig.\ref{large}(a) \citep{Urquhart2018-473,Urquhart2022-510}, they well correspond to different velocity components shown in the moment-1 maps of $^{13}$CO (3$-$2) (Fig.\ref{large}(b)). peak1 and peak2 associate with the clump clusters C and D. Both A and B clump clusters are located in the velocity component peak3. 
peak1 is relatively independent of peak2 and peak3, but the dominant features in peak2 and peak3 appear have some overlap, yet they can be separated. Hence, we fit Gaussian functions in order to minimise the effects of cross-contamination. 
peak2 and peak3 can be distinguished by the full width at half maximum (FWHM).
In this paper, we mainly focus on peak3 with a velocity range [$-$60, $-$35]\,km\,s$^{-1}$, which represents a complete shell structure shown in Fig.\,\ref{large}(e). The association between peak1, peak2 and peak3 will be discussed in detail in a forthcoming paper. 

In order to study the structure of the molecular cloud complex in position-position-velocity space,  we applied the  fully automated Gaussian decomposer \texttt{GAUSSPY+} \citep{Lindner2015L,Riener2019-628}, which can decompose the complex spectra of molecular lines into multiple Gaussian components. The most important four parameters in the algorithm are the first and second smoothing parameters $\alpha_1$ and $\alpha_2$, signal-to-noise (SNR) and significance criterion. To determine the smoothing parameters $\alpha_1$ and $\alpha_2$, sets with 500 randomly selected spectra from the $^{13}$CO (3$-$2) cube were used to train \texttt{GAUSSPY+}.
The default values of SNR and significance are 3 and 5, but considering that the moment-1 map needs a 5$\sigma$ threshold to show a clear velocity field, we adjust the value of SNR to 5. Other parameters for the decomposition are set to the default values provided by \citet{Riener2019-628}.

As shown in Fig.\,\ref{ppv},
the velocity components separated by the average spectrum is consistent with the decomposition of \texttt{GAUSSPY+}, with the three main velocity components well separated in PPV space. 
Different from peak1 and peak2, for peak3, no matter the projection on Galactic longitude or latitude, we can see a large-scale continuous structure, indicating the structures in peak3 are physically connected. 
Especially, the overall structure of peak3 consists of a large-scale shell structure extended from $l=330.6\degr$ to $l=334\degr$. It may be generated by a large-scale compression, such as the shock from supernova explosion events, which will be investigated in a forthcoming paper. 

The entire G333 complex seems to have fragmented into several independent molecular clouds, they are treated as sub-regions marked in Fig.\,\ref{sub}, such as s3. Sub-structures in each sub-region marked by letters in Fig.\,\ref{sub}(a) are divided by the bright emission of $^{13}$CO (3-2) shown in red contours, such as s3a.  
From Fig.\,\ref{sub}(b) and Fig.\,\ref{large}(a), we can see the dense parts of $^{13}$CO (3$-$2) emission tightly correlate with the infrared bright regions. The overlap of different velocity components in sub-regions 5 and 7 is significant, as shown in Fig.\ref{large}(a). However, 
these two sub-regions only include a few structures in the velocity range [$-$60, $-$35]\,km\,s$^{-1}$, thus they are subordinate in G333 complex. 

Through the above analysis, we see that the G331  GMC and the G333 complex overlap on the sky, but they have different velocities and lie at very different distances. Given that the G333 region is closer to us, we focus the remainder of the paper on it.

\subsection{Identification of filaments}

The first step is to identify and characterize filaments in the G333 complex.
Following the method described in \citet{Zhou2022-514}, we use the Moment 0 maps (the integrated intensity maps) of $^{13}$CO (3$-$2) to identify filaments in the G333 complex using the FILFINDER algorithm \citep{Koch2015}. The velocity intervals for making Moment 0 maps are determined from the averaged spectra of each sub-region, 
marked by vertical dotted lines in Fig.\,\ref{sspectra}. Moreover, a threshold of 5$\sigma$ is applied to make the moment maps, which can reduce the noise contamination effectively.  
The skeletons of identified filaments overlaid on the moment-0 maps of $^{13}$CO (3-2) line emission are shown in Fig.\,~\ref{fil}. They are highly consistent with the gas structures traced by $^{13}$CO (3-2) as seen by eye, indicating that the structures identified in FILFINDER are reliable. Then for each sub-region, we break the filamentary network into several filaments. This step is necessary to calculate the offset along the filament before fitting the velocity gradient. 
To show the velocity field along the filament more clearly, we try to make each filament long enough. 

Considering that each sub-region includes many sub-structures, it is not surprising that their average spectra will show multiple velocity peaks. In Fig.\,\ref{sspectra}, the average spectra of G333-s3 clearly shows multiple components. We separate out each peak in the line profile, and find that each peak indeed corresponds to a sub-structure. This situation may also exist in other sub-regions. However, generally these sub-structures are well spatially separated and thus do not have a significant impact on the moment maps analysis.

Fig.\,\ref{hub} displays the \texttt{GAUSSPY+} decomposition of each sub-region in PPV space. From there, we can see the main structures of each sub-region are connected in PPV space and thus unlikely to be contaminated by velocity components from unrelated foreground or background cloud emission. Moreover, after we extract the velocity and intensity information along the filament skeletons, if there is overlap of unrelated velocity components, we should find anomalies in Fig.\,\ref{example}, such as a sudden break at a certain position of the skeleton. However, that is not the case in our analysis.
Furthermore, we only analyze the variations of velocity and intensity along the filament skeletons with a one pixel width, which can also effectively avoid potential overlap. In \citet{Zhou2022-514}, we found that multiple velocity components are common in hub-filament systems, especially in hub regions. Furthermore, 
the observed samples in \citet{Zhou2022-514} with or without multiple velocity components show similar results, such as the velocity gradient estimated from moment maps.

To say the least, we can assume that most of the pixels above the filaments have a single velocity component or have a dominant velocity component, and then we compare the fitting results with the free-fall model and previous results from ALMA data, if they are consistent, indicating that the hypothesis is reasonable. That is the case in our work. 
Conversely, if the overlap of velocity components is significant, the fitted velocity gradient should be free without the regularities shown in our work, because the overlap of unrelated velocity components is random. 
From Fig.\,\ref{example}, we can see good correlations between velocity and density fluctuations, which is consistent with previous studies \citet{Hacar2011-533,Henshaw2014-440,Henshaw2016-463,Liu2019-487,Henshaw2020-4}. The variation of the velocity gradient with the scale shown in Fig.\,\ref{gradient} is also consistent with the results in \citet{Zhou2022-514}.

\subsection{Velocity gradients}\label{vg}

The kinematical features of hub-filament systems can be revealed by the velocity gradients along the filaments. However, it is difficult to directly carry out the measurement in PPV space due to the complex gas motion shown in Fig.\,\ref{ppv}. 
In \citet{Zhou2022-514},
we investigate the presence of hub-filament systems in a large sample of 146 proto-clusters using H$^{13}$CO$^{+}$ J=1-0 molecular line in ATOMS survey. The strongest intensity peaks of H$^{13}$CO$^+$ emission coincide with the brightest 3 mm cores or hub regions in those hub-filament systems. We find that filaments are  ubiquitous in proto-clusters, velocity and density fluctuations along these filaments are seen. We firstly estimated two overall velocity gradients between velocity peaks and valleys at the two sides of the center of the gravitational potential well, and ignored local velocity fluctuation. We also derived additional velocity gradients over smaller distances around the strongest intensity peaks of H$^{13}$CO$^+$ emission (see Fig.\,6 in \citealt{Zhou2022-514}). 
In this work, the same method is used to derive the velocity gradients from the LAsMA data. Then the fitted velocity gradients derived from the LAsMA and ALMA data are combined for a statistical analysis, as shown in Fig.~\ref{gradient}.

Fig.\,\ref{example} shows the intensity-weighted velocity (Moment 1) and integrated intensity (Moment 0) of $^{13}$CO (3$-$2) line emission along the skeletons of selected filaments with intense
velocity and density fluctuations along these filaments.
Generally, the peaks of the density fluctuations are associated with ATLASGAL clumps as shown in Fig.\,\ref{fil}. The density and velocity fluctuations along filaments may indicate converging gas flows coupled to regularly spaced density enhancements that probably form via gravitational instabilities \citep{Henshaw2020-4}.
In Fig.\,\ref{example}, we can see the typical V-shape velocity structure around the intensity peaks, which we fit as velocity gradients on both sides of the intensity peak marked by the straight lines. Fig.\,\ref{example}(a) shows a more complex velocity field with more repetitive segments than the simple structure in Fig.\,\ref{example}(b). Gas kinematics of other long filaments in Fig.\,\ref{fil} are similar to Fig.\,\ref{example}(a). For the long filaments, except for the local velocity fluctuations or gradients, they also exhibit velocity gradients on larger scales, which are fitted by ignoring the local velocity fluctuations. Generally, large-scale velocity gradients are associated with several intensity peaks.

\subsection{Hub-filament structure}\label{hub1}

Each sub-region marked in Fig.\,\ref{sub}(a) contains many sub-structures, which are nearly separated from each other, except for several crowded regions. Red contours from  $\sim$50\% of the integrated intensity of the $^{13}$CO (3-2) emission show the high density parts of various structures.  
In Fig.\,\ref{sub}, several typical hub-filament structures are highlighted in the first row, see also Fig.\,\ref{fil} for more details. We can see filamentary structures connected to high-density hubs. Other sub-structures in the entire observed region have more or less similar morphology. Some differences are expected, since the appearance of hub-filament morphology relies on the projection angle and the resolution of observation. 
Moreover, most structures have the typical kinematic features of hub-filament systems, as discussed in next sections. For example, the peaks of intensity (the hubs) are associated with the converging velocities, 
indicating that the surrounding gas flows may be converging to dense structures. Especially, velocity gradients increase at small scales.


Orange circle in Fig.\,\ref{example} marks the intensity peak of $^{13}$CO (3$-$2) emission associated with an ATLASGAL clump where the distance between the central coordinate of ATLASGAL clump and the intensity peak of $^{13}$CO (3$-$2) emission is less than the effective radius of ATLASGAL clump. 
These associated clumps are treated as hubs in this work, their properties are listed in Table\,\ref{clump}. In the second part of Table \,\ref{clump}, some of the peaks have more than one corresponding ATLASGAL clumps and we have added those together, as indicated by a space in the table. We focus on their mass, effective radius and the corresponding velocity gradients. However, as shown in Fig.\,\ref{example},
there are also many peaks of $^{13}$CO (3$-$2) emission without corresponding clumps, thus not included in the table, but they are similar to the peaks associated with clumps. Generally, large-scale velocity gradients involve many intensity peaks with or without corresponding clumps, thus it is difficult to estimate the properties of their hubs, therefore these are not given in the table.



\subsection{Funnel structure}\label{funnel}

If we compare the velocity gradients of molecular gas to iron filings, just as the iron filings can reflect the shape of a magnetic field, the distribution of velocity gradients may to some extent reflect the shape of the force field that plays the dominant role in the molecular clouds. Below, we will investigate the distribution of the molecular velocity gradients in PPV space. 
In the case of gravitational free-fall onto a central point mass, velocities change as $v \propto r^{-1/2}$ ($r = \sqrt{x^{2}+y^{2}}$). In the case of collapse driven by the gravity of a gaseous mass with a certain density profile, the radial dependence of the infall speed depends on the slope of the density profile \citep{Gomez2021-502}. In either case, this results in a ‘funnel’ morphology of the velocity field in PPV space, as illustrated in Fig.\,\ref{model}.
The exterior cloud velocities show only gentle variations (small velocity gradient, funneling material from cloud to clump), while the velocities in the interior part near the center change dramatically with scales (large velocity gradient, funneling material from clump to core). 
We note that
the schematic in Fig.\,\ref{model} is oversimplified. 
Generally a molecular cloud contains many clumps inside, and a clump also contains many cores. Here, we choose only one clump and one core to demonstrate the gas inflow and the variation of the velocity gradient from the molecular cloud scale to the core scale. A more realistic case is shown in Fig.\,\ref{hub}, here we indeed find 
most of sub-structures show the expected funnel structures in PPV space. Particularly, close to the intensity peak, the velocity gradient increases steeply, also reflected in Fig.\,\ref{example}. 

Moreover, the V-shape velocity structure around the intensity peaks indicates accelerated material inflowing towards the central hub \citep{Gomez2014-791, Kuznetsova2015-815, Hacar2017-602, Kuznetsova2018-473, Zhou2022-514}, which can be seen in Fig.\,\ref{example} everywhere. The V-shape velocity structure can be treated as the projection of the funnel structure from  PPV space to PV plane. Hence, the funnel structures in PPV space can be an effective probe for the gravitational collapse motions in molecular cloud.


\section{Discussion}


\subsection{Characteristic scale $\sim$1\,pc}

In this work, 
we have investigated gas motions on sufficiently large scales, the longest filament $\sim$50\,pc, obtaining similar results to previous small-scale ALMA observations. 
We fitted a wide range of values of the velocity gradients, as shown in Fig.\,\ref{gradient}(a). The velocity gradients are small at large scales, while they become significantly larger at small scales ($\lesssim 1$ pc), as is the case in the ATOMS survey \citep{Zhou2022-514}.  
For both ALMA and LAsMA data, most of the fitted velocity gradients concentrate on the
scales $\sim 1$\,pc, as shown in Fig.\,\ref{gradient}. This is consistent with the fact that 1\,pc is considered as the characteristic scale of massive clumps \citep{Urquhart2018-473}. In Fig.\,\ref{gradient}, the variation of  velocity gradients with scales is comparable with the expectations from gravitational free-fall, in the sense that the gradient decreases smoothly with increasing scale.
Fig.\,\ref{mas}(c) and (d) show similar results to Fig.\,\ref{gradient}, but only consider the velocity gradients associated with ATLASGAL clumps, listed in Table\,\ref{clump}. Especially, in Fig.\,\ref{mas}(c), the required central mass in the fitted free-fall model is consistent with the mass distribution observed in Fig.\,\ref{mas}(b).

\subsection{Evidence for Gravitational Acceleration}\label{acceleration}

Fig.\,\ref{mas}(a) displays the probability distribution of velocity gradients measured around 1\,pc (0.8$\sim$1.2 pc), showing that the most frequent velocity gradient is $\sim 1.6$\,km\,s$^{-1}$\,pc$^{-1}$. Assuming free-fall, 
\begin{equation}
\nabla V_{free}= -\frac{d}{dR}\sqrt{\frac{2GM}{R}}
=\sqrt{\frac{GM}{2R^3}},
\label{v-free}
\end{equation}
we estimate the kinematic mass corresponding to 1\,pc is $\sim1190$\,M$_\odot$, which is comparable with the typical mass of clumps in the ATLASGAL survey \citep{Urquhart2018-473}. In Fig.\,\ref{mas}(b), the peak of the associated clump's mass distribution is also around $\sim$ 1000 M$_\odot$. Thus the clump may be a gravity-dominated collapsing object, also consistent with the survey results that most Galactic parsec-scale massive clumps seem to be gravitationally bound no matter how evolved they are \citep{Liu2016-829, Urquhart2018-473, Evans2021-920}. 

Based on Eqn.\,\ref{v-free}, we can use the values of $M$, $L$ and $\nabla V$ listed in Table\,\ref{clump} to compare observations to the simple theory. In Fig.\,\ref{free}, the observed and predicted accelerations do show a clear correlation, strong evidence for gravity accelerating the gas inflow. 
The scatter in Fig.\,\ref{free} is to be expected because we measure only the projection of the realistic velocity vector and hence acceleration. For the same reason, we tend to underestimate the observed acceleration. 
Moreover, in Table\,\ref{clump}, we only selected the ATLASGAL clumps associated with the intensity peaks of the $^{13}$CO (3$-$2) emission (a deviation within the effective radius of the clump), thus some of them are not the exact gravitational centers. Despite these caveats, the correlation between observations and predictions for the denser clumps is quite good. The less dense clumps tend to lie {\it above} the line of equality, suggesting their masses underestimate the total mass contributing to the gravitational acceleration.

\subsection{Large-scale inflow driven by gravity} \label{picture}

 As shown in Fig.\,\ref{gradient} and Fig.\,\ref{mas}, the velocity gradients fitted to the LAsMA and ALMA data agree with each other over the range of scales covered by ALMA observations in the ATOMS survey ($\textless$ 5\,pc). Interestingly, the variations of velocity gradients at small scales ($\textless$ 1\,pc), middle scales ($\sim$ 1-7.5\,pc) and large scales ($\textgreater$ 7.5\,pc) are consistent with gravitational free-fall with central masses of $\sim$ 500\,M$_\odot$, $\sim$ 5000\,M$_\odot$ and $\sim$\, 50000\,M$_\odot$. It means the velocity gradients on larger scales require larger mass to maintain, which is also revealed by the funnel structure shown in Fig.\,\ref{model}. Indeed, larger masses imply larger scales, that is to say, the larger scale inflow is driven by the larger scale structure which may be the gravitational clustering of smaller scale structures (Sec.\,\ref{order}), consistent with the hierarchical structure in molecular clouds and the gas inflow from large to small scales.
Thus the large scale gas inflow may also be driven by gravity, with the molecular clouds in G333 in the state of globally gravitational collapse, which is consistent with the argument that these molecular clouds act as cloud-scale hub-filament structures, see the description in Sec.\,\ref{hub1}. Moreover, as described in Sec.\,\ref{funnel}, the funnel structure of the velocity field shown in PPV space also tends to support the large-scale gravitational collapse of the molecular clouds in G333 complex.


\subsection{Hierarchical gravitational structure in molecular cloud}\label{order}

The scales of hub-filament structures (cloud-scale) in this work are much larger than the ones (clump-scale) in \citet{Zhou2022-514}. Each large-scale hub-filament structure will contain many sub-structures, thus the kinematic structure will be very complex. 
Nevertheless,
if they are hub-filament systems, we expect that they have similar kinematic features like the clump-scale hub-filament system. A hub-filament system must have a common gravitational center, which will dominate the overall gravitational field, thus the velocity field. 
In Fig.\,\ref{ppv}, although there are many sub-structures in each sub-region, which cause many velocity and density fluctuations, globally, the entire large-scale structure still displays the funnel structure.  As discussed in Sec.\,\ref{vg} and Sec.\,\ref{picture}, large-scale velocity gradients always involve many intensity peaks, and the larger scale inflow is driven by the larger scale structure, implying the clustering of local small-scale gravitational structures can act as the gravitational center on larger scale. To some extent, the funnel structure gives an indication of the gravitational potential well formed by the clustering.

Both \citet{Zhou2022-514} and this work find that when the scale is less than $\sim$1\,pc, 
the gravitationally dominated gas infall can become apparent. Thus, relatively small-scale hub–filament structures will have better and more recognizable morphology than large-scale ones due to the strong local gravitational field. For the large-scale hub–filament structures, background turbulence is more likely to disturb their morphology due to weaker gravitational confinement on large scales, thus they are more "blurred". 
\citet{Zhou2023-519} suggested for the interpretation of the formation of the large-scale hub-filament structure in W33 complex, that IRDCs and ATLASGAL clumps in W33-blue can be treated as small-scale hub–filament structures, W33-blue itself as a huge hub–filament structure. Thus there may be a self-organization process from small-scale hub–filament structures to a large-scale one. If the large scale gravitational center is the clustering of small scale ones, compared to the latter, large scale gravity center will be more loose, thus its ability to control the gas motions on large scale is also weaker, reflected as the loose funnel structure and small velocity gradient, as shown in this work. In Fig.\,\ref{hub}, the small-scale structures indeed have clearer funnel structures than the large-scale ones, implying the role of gravity in shaping the funnel structure.

\subsection{G333 in Context of Galactic Molecular Clouds}\label{context}

The same region of the sky, with slightly larger area ($332.6\degr<l<333.8\degr$ and $-0.8\degr<b<0.1\degr$),  has also been studied by \citet{2015ApJ...812....7N},
using CO and $^{13}$CO (1-0) emission and archival data. They call the G333 GMC the RCW 106 complex after the associated HII region. They determined the following properties for the RCW 106 complex: the total effective diameter is 183 pc, the mass is 5.9\ee6\,\msun, and the mass surface density is 220\,\msun\,pc$^{-2}$. The virial parameter for the whole complex is 0.35, making it strongly gravitationally bound; since the cloud is elongated, they include a linear analysis that also implies a strongly bound cloud. These properties would make the G333 GMC stand out as one of the largest, most massive complexes in the Galaxy (cf Fig.\,7 of \citealt{Miville2017-834}), and one of the few that is gravitationally bound  (cf Fig.\,2 of \citealt{Evans2021-920}). However, \citet{2015ApJ...812....7N} took the velocity range of the RCW 106 complex as [-80, -40] km s$^{-1}$. As discussed in Sec.\,\ref{components}, after a careful decomposition of velocity components, we find the velocity range [-80, -40] km s$^{-1}$ also includes G331 GMC; thus the mass of the G333 GMC is overestimated in 
\citet{2015ApJ...812....7N}.  The catalog of \citealt{Miville2017-834}, restricted to the velocity range of peak3, yields a mass for the G333 cloud traced by CO 1-0 emission of $\sim$ 1.7 $\times$ 10$^{6}$ M$_{\odot}$, with the surface density  $\sim$ 120\,\msun\,pc$^{-2}$.
Tracers that are biased toward higher volume densities find progressively smaller masses. 
\citet{Karnik2001-326} made a far-infrared (FIR) 150 and 210\,$\mu$m dust emission study of the region, using the canonical gas:dust ratio of 100, they measure a total GMC mass of $\sim$ 1.8\,$\times$ 10$^{5}$\,M$_{\odot}$. This is consistent with the total mass of $\sim$ 1 $\times$ 10$^{5}$\,M$_{\odot}$ estimated by \citet{Mookerjea2004-426} from a 1.2-mm cold dust continuum emission map and the same gas-to-dust ratio. These estimates show the decreasing fraction of mass in progressively denser regions.

The lower total mass from CO 1-0 based on the catalog of \citet{Miville2017-834} still makes the G333 complex one of the most massive in the Milky Way. Analysis of the RCW HII region yields a stellar mass of 48\ee3\,\msun; assuming a timescale of 0.2\,Myr (the lifetime of an O7 star), \citet{2015ApJ...812....7N} derive a star formation rate so far of 0.25\,\msun\,Myr$^{-1}$ and an efficiency ($M_*/M_{\rm gas}$) of 0.008. 
The very high star formation rate led \citet{2015ApJ...812....7N} to describe it as a "mini-starburst". It is thus atypical of most molecular clouds but an exemplar of the very small subset that form the majority of stars in the Galaxy.

\section{Summary}
We investigated the gas kinematics of hub-filament structures in the G333 complex using $^{13}$CO (3$-$2) emission. The main conclusions are as follows:\\

1. The G333 complex includes three main velocity components, which can be well separated by the average spectra and the velocity distribution in PPV space. Here the  velocity distribution is derived from the decomposition of $^{13}$CO (3-2) emission by \texttt{GAUSSPY+}. The velocity components with velocity ranges [$-$120, $-$76]\,km\,s$^{-1}$ and  [$-$76, $-$60]\,km\,s$^{-1}$ represent the main two parts of G331 GMC, the third velocity component with velocity range [$-$60, $-$35]\,km\,s$^{-1}$ shows a very extended shell structure from $l=330.6\degr$ to $l=334\degr$ with G333 GMC and G332 ring as the main parts. \\

2. The entire G333 complex seems to have fragmented into several independent molecular clouds, with some of the sub-structures in these molecular clouds showing typical hub-filament structures, which also have the typical kinematic features of hub-filament systems. The broken morphology of some very infrared bright structures indicates that the feedback is disrupting the star-forming regions.\\

3. We use the integrated intensity maps of $^{13}$CO J=3-2 to identify filaments in G333 complex by FILFINDER algorithm, and extract the velocity and intensity along the filament skeleton from moment maps. We find a good correlation between velocity and density fluctuations, and fit the velocity gradients around the intensity peaks.
The change in velocity gradients with the scale indicates that the morphology of the velocity field in PPV space resembles a ``funnel'' structure.
The funnel structure can be explained as accelerated material inflowing towards the central hub and gravitational contraction of star-forming clouds/clumps. The typical V-shape velocity structure along the filament skeleton can be treated as the projection of the funnel structure from PPV space to PV plane. Hence, the funnel structure in PPV space can be an effective probe for the gravitational collapse motion in molecular cloud.\\

4. We have investigated gas motions on sufficiently large scales, the longest filament $\sim$50\,pc, but obtain similar results to small-scale ALMA observations. 
The typical velocity gradient corresponding to a one pc scale is $\sim 1.6$\,km\,s$^{-1}$\,pc$^{-1}$. Assuming the free-fall model, we can predict the gravitational acceleration onto each hub. The observed accelerations correlate well with the predicted ones and have values comparable to the prediction. Given the fact that we observe only one component of the acceleration and that the masses of the hubs are uncertain, the result provides strong evidence for gravitational acceleration of material flowing into hubs from filaments.

5. The velocity gradients fitted by LAsMA data and ALMA data agree with each other over the scales covered by ALMA observations in the ATOMS survey ($\textless$ 5\,pc). On large scales, we find that the larger scale inflow is driven by the larger scale structure, indicating the hierarchical structure in molecular cloud and the gas inflow from large scale to small scale. The large scale gas inflow is driven by gravity, implying that the molecular clouds in G333 may be in the state of globally gravitational collapse. The funnel structure of velocity field shown in PPV space also tends to support the large-scale gravitational collapse.\\

6. Although there are many sub-structures in each molecular cloud, which cause ubiquitous velocity and density fluctuations, totally, the entire large-scale structure still displays the loose funnel structure. Thus the hub-filament structures at different scales may be organized into a hierarchical macroscopic system through the coupling of gravitational centers at different scales. \\

In short, changes of velocity gradient with scale indicate a ”funnel” structure of the velocity field in PPV space, indicative of a smooth, continuously increasing velocity gradient from large to small scales, and thus consistent with gravitational acceleration.



\begin{acknowledgements}
This publication is based on data acquired with the Atacama Pathfinder Experiment (APEX) under programme ID M-0109.F-9514A-2022. APEX has been a collaboration between the Max-Planck-Institut für Radioastronomie, the European Southern Observatory, and the Onsala Space Observatory. This work has been supported by the National Key R\&D Program of China (No. 2022YFA1603100). Tie Liu acknowledges the supports by National Natural Science Foundation of China (NSFC) through grants No.12122307 and No.12073061, the international partnership program of Chinese Academy of Sciences through grant No.114231KYSB20200009, and Shanghai Pujiang Program 20PJ1415500.
\end{acknowledgements}

\bibliographystyle{aa} 
\bibliography{G333} 

\begin{appendix}
\twocolumn

\section{Supplementary maps}\label{supply}

\begin{figure*}
\centering
\includegraphics[width=0.9\textwidth]{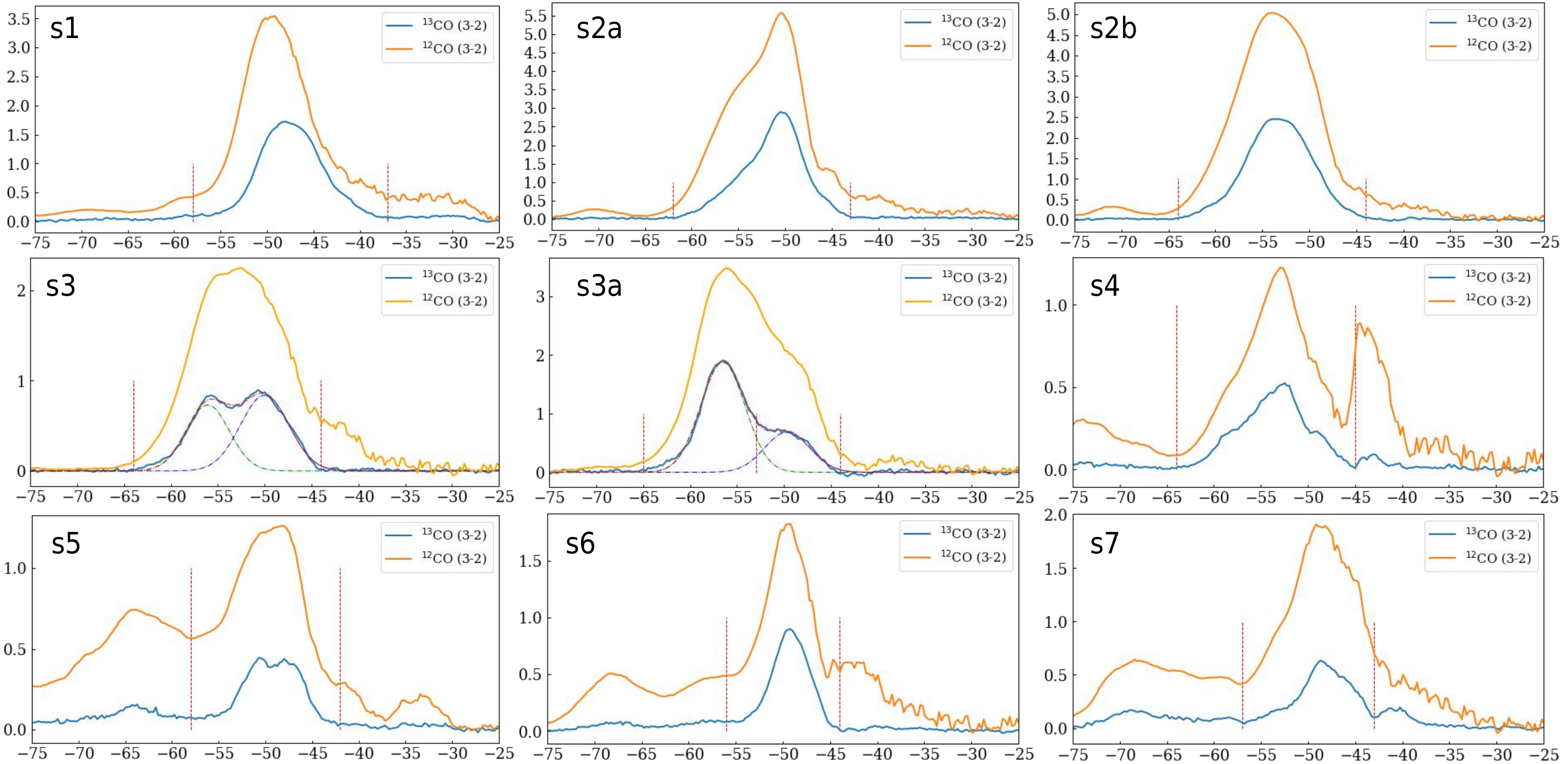}
\caption{Average spectra of $^{13}$CO (3-2) and $^{12}$CO (3-2) for the sub-regions shown in Fig.\ref{fil}. Vertical dashed lines mark the velocity ranges which are used to make moment maps.}
\label{sspectra}
\end{figure*}

\begin{figure*}
\centering
\includegraphics[width=0.8\textwidth]{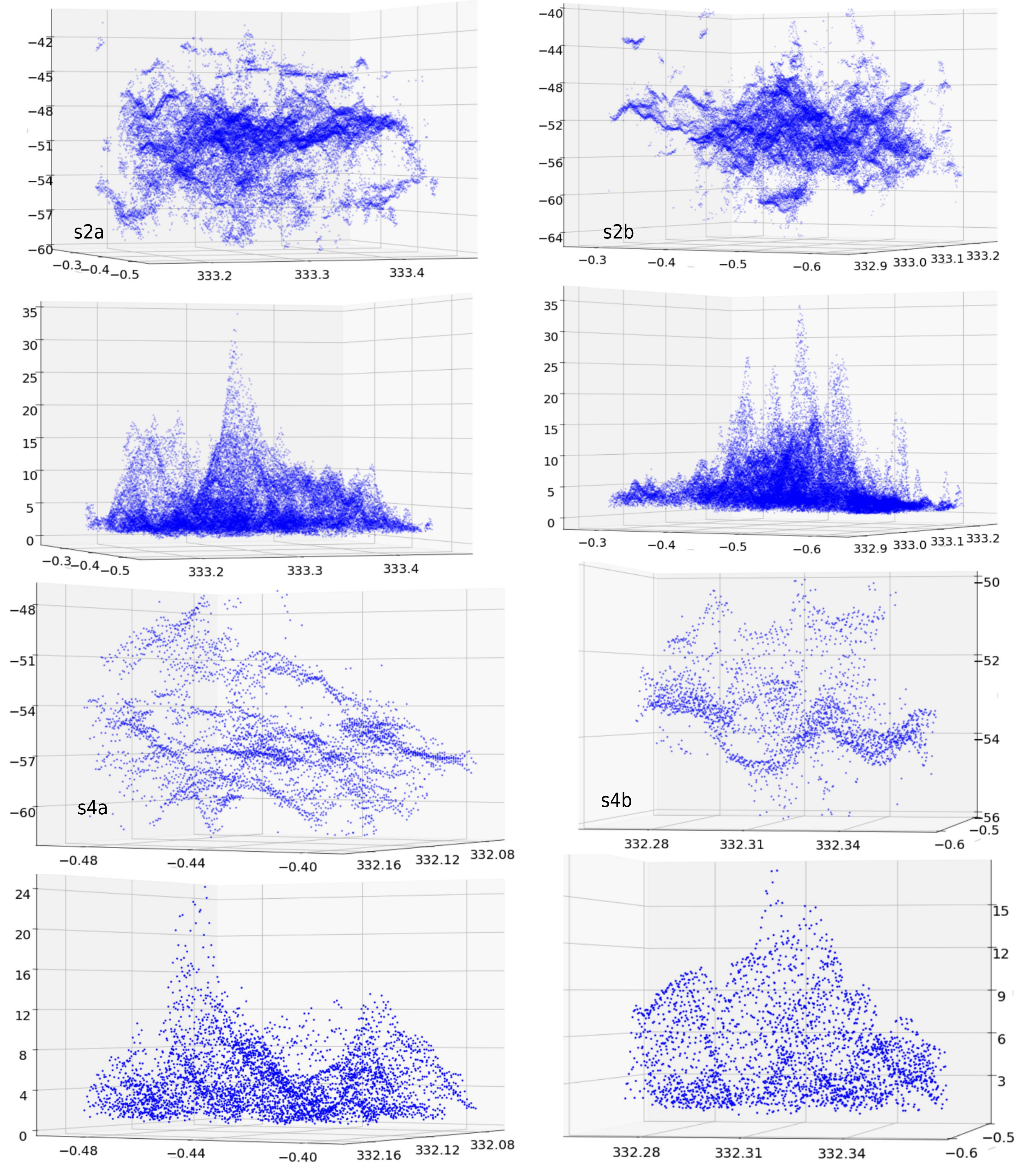}
\caption{The same with Fig.\ref{hub}.}
\label{hub2}
\end{figure*}
\begin{figure*}
\centering
\includegraphics[width=0.8\textwidth]{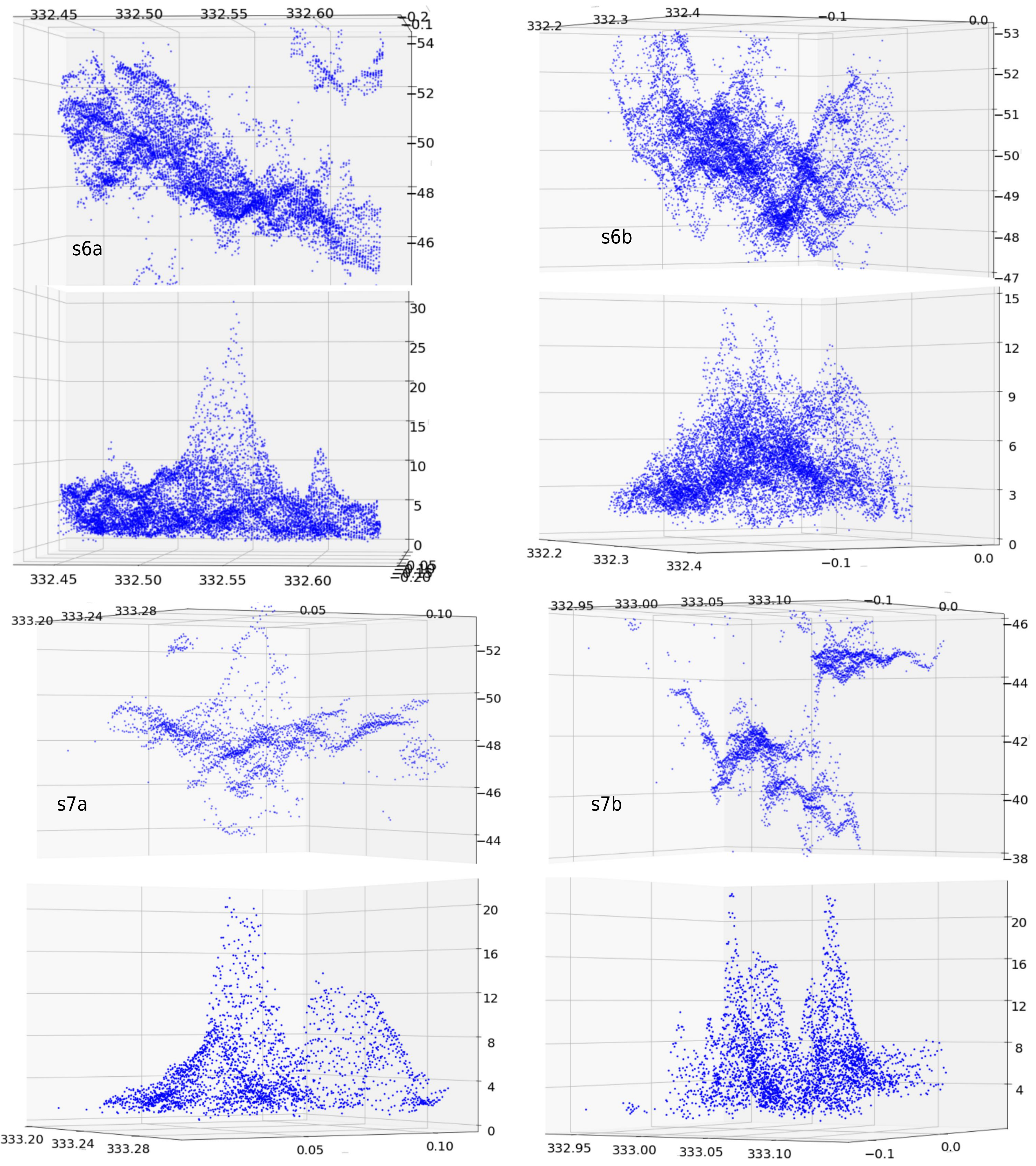}
\caption{The same with Fig.\ref{hub}.}
\label{hub3}
\end{figure*}

\begin{figure*}
\centering
\includegraphics[width=1\textwidth]{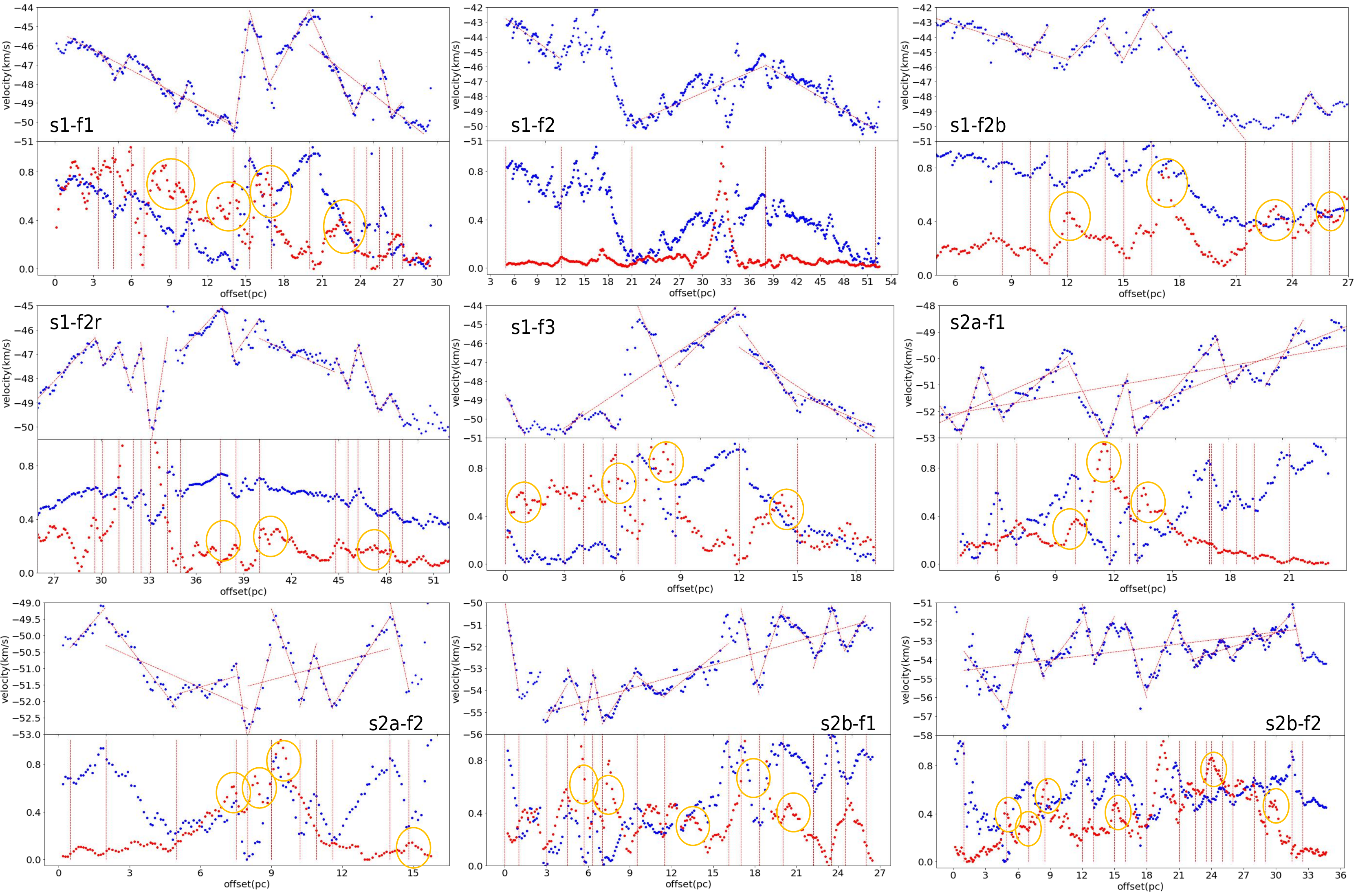}
\caption{The same with Fig.\ref{example}.
}
\label{example1}
\end{figure*}

\begin{figure*}
\centering
\includegraphics[width=1\textwidth]{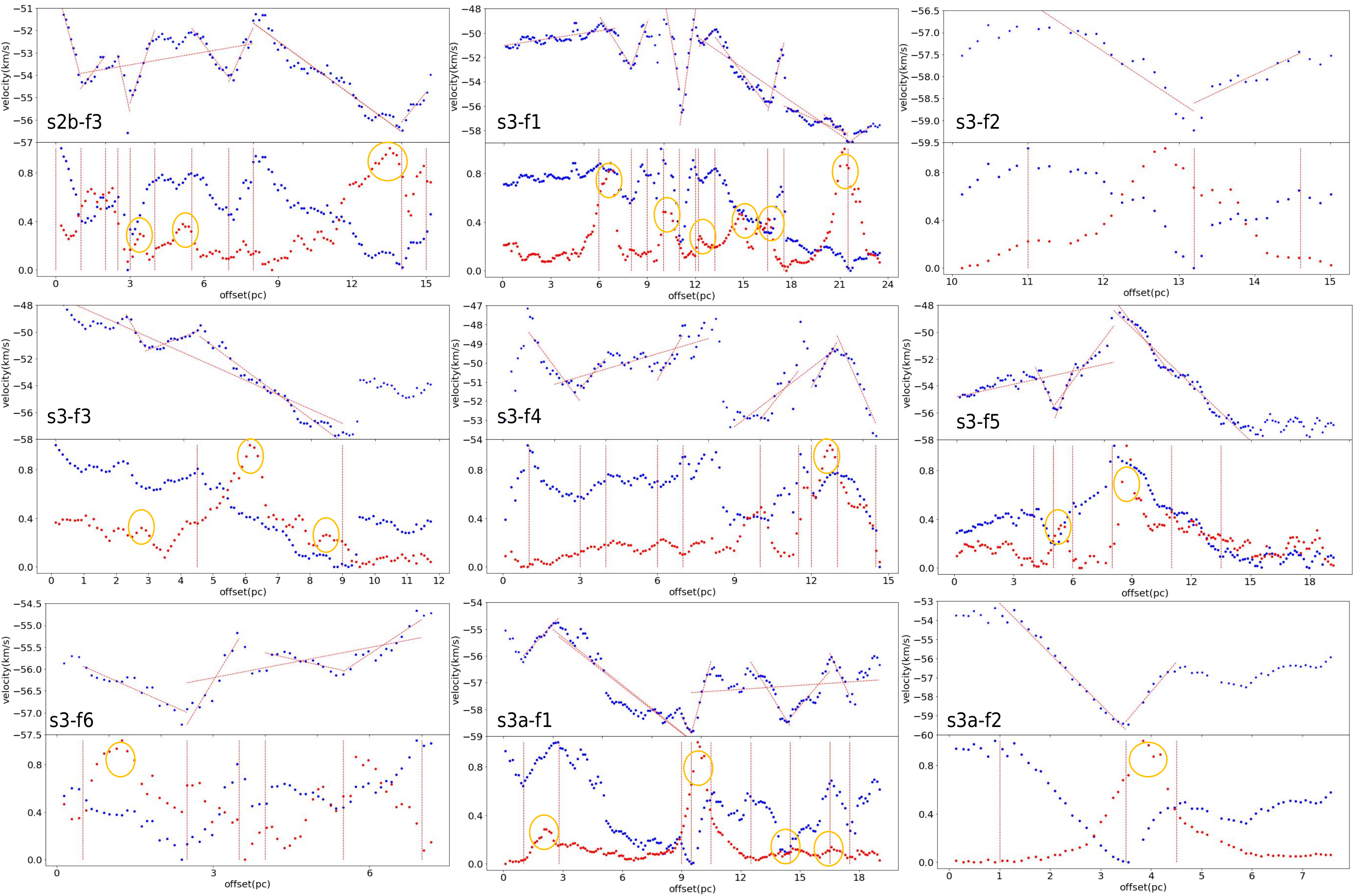}
\caption{The same with Fig.\ref{example}.}
\label{example2}
\end{figure*}
\begin{figure*}
\includegraphics[width=1\textwidth]{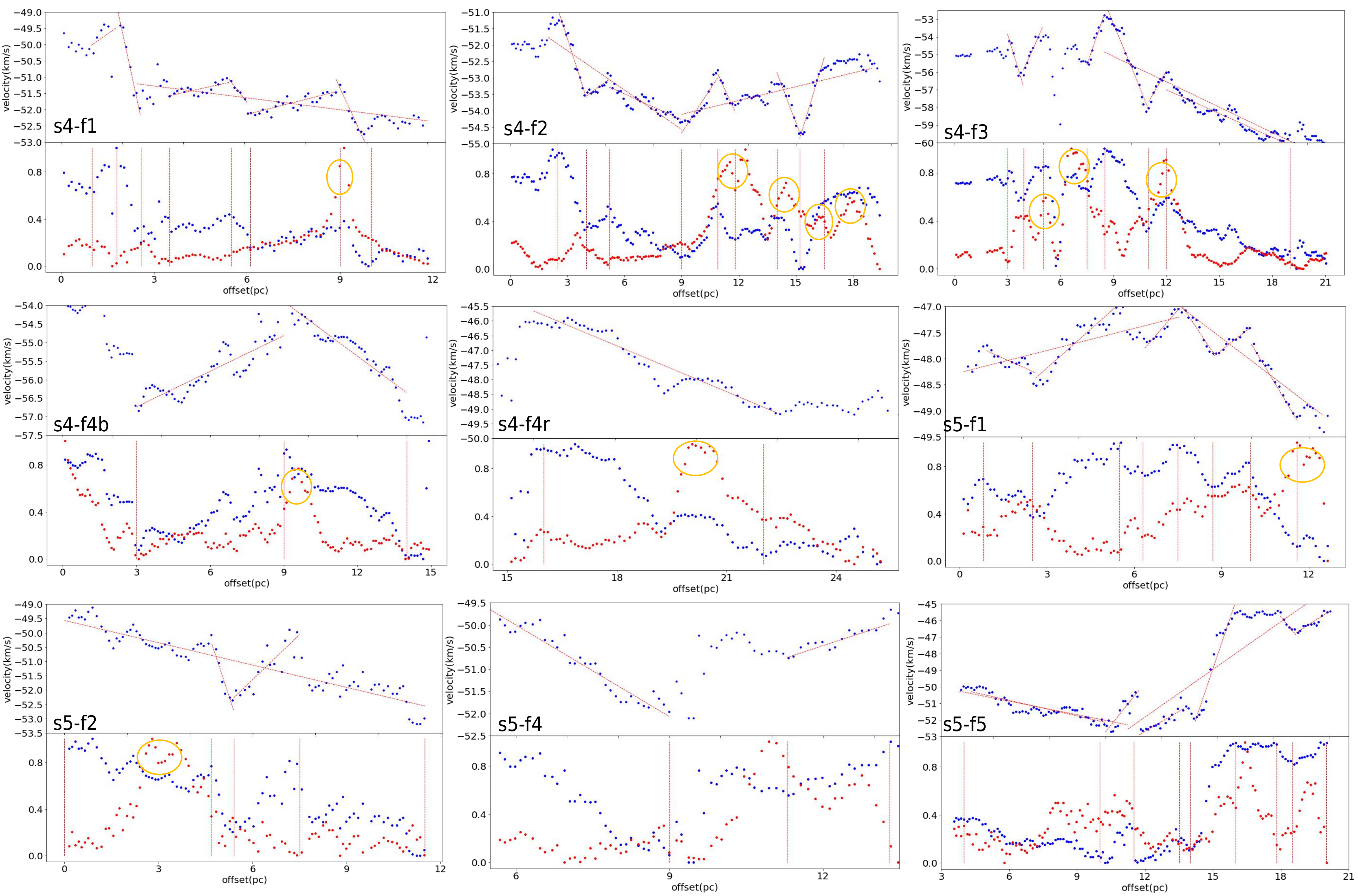}
\caption{The same with Fig.\ref{example}.}
\label{example3}
\end{figure*}
\begin{figure*}
\centering
\includegraphics[width=1\textwidth]{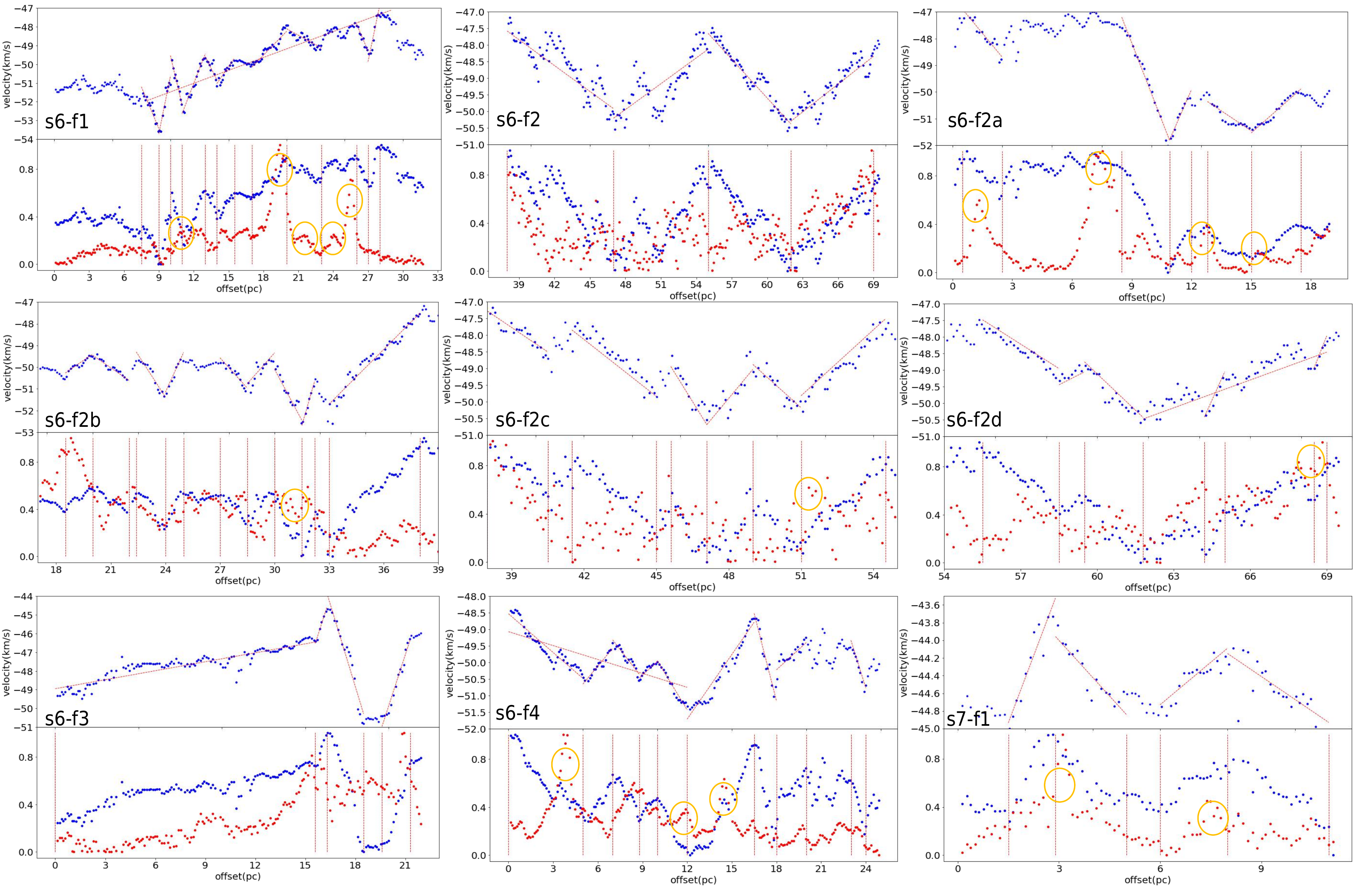}
\caption{The same with Fig.\ref{example}.}
\label{example4}
\end{figure*}
\begin{figure*}
\centering
\includegraphics[width=1\textwidth]{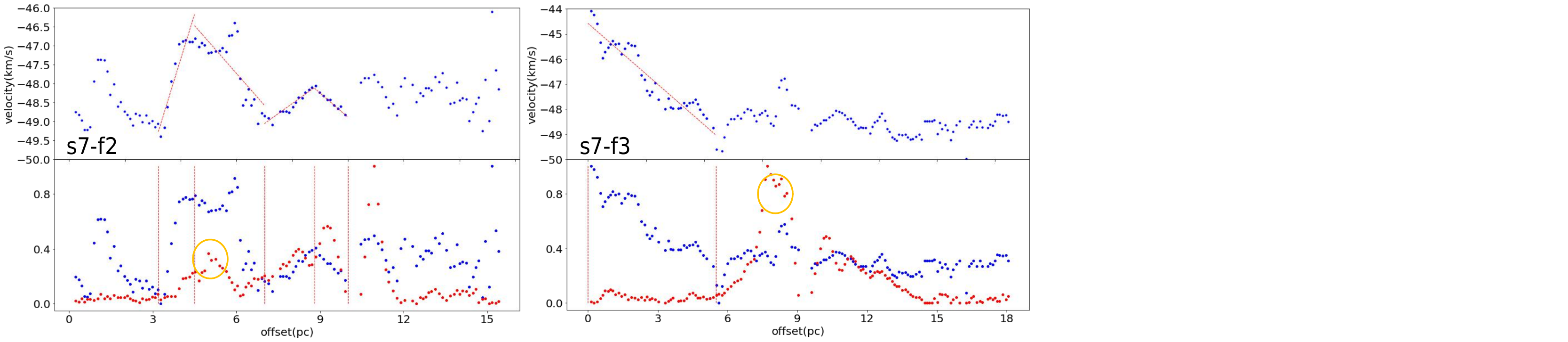}
\caption{The same with Fig.\ref{example}.}
\label{example5}
\end{figure*}

\section{Appendix A:  Noise and offset details}\label{app-a}

\begin{figure*}
\centering
\includegraphics[width=0.9\textwidth]{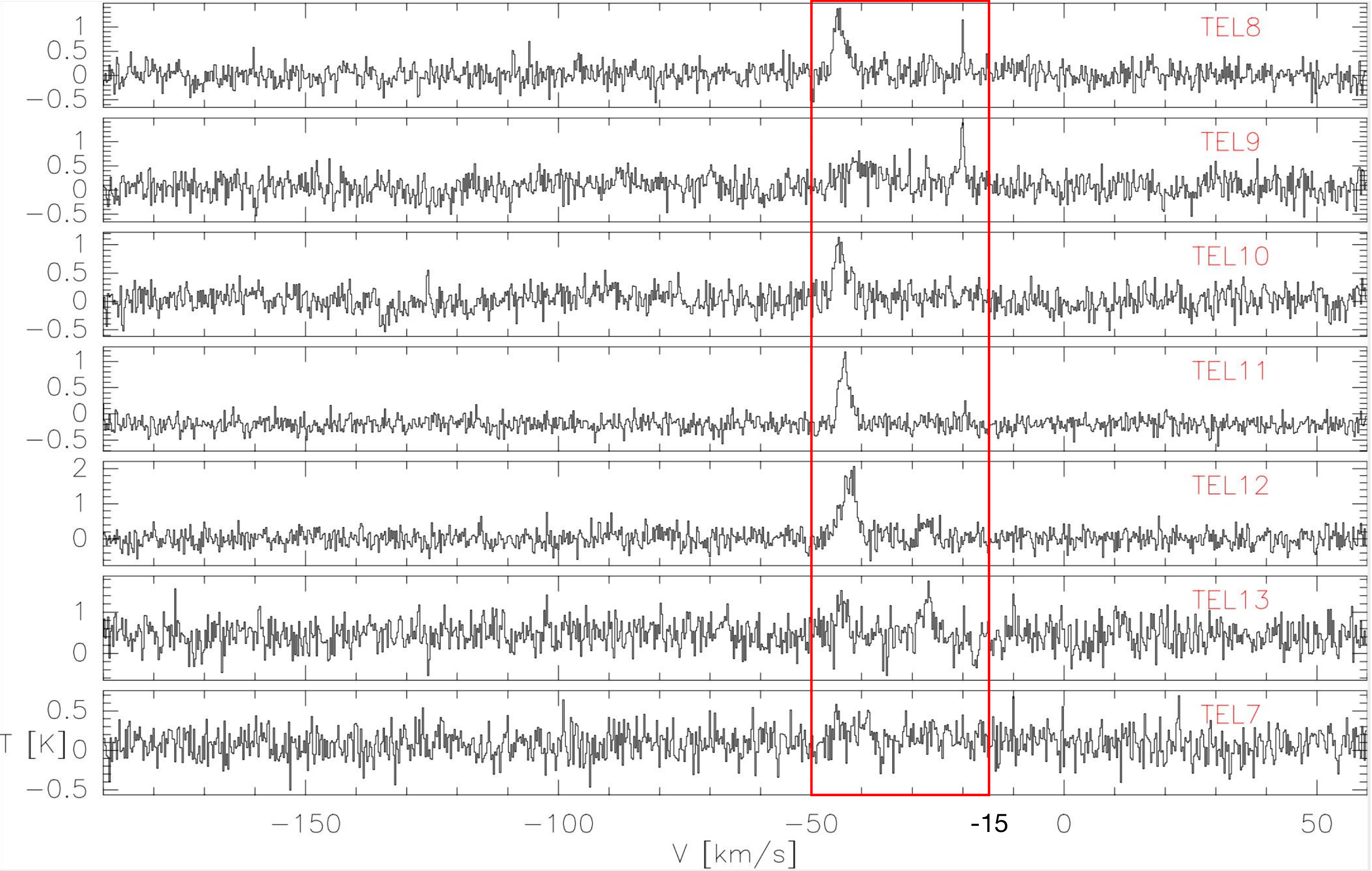}
\caption{Average spectra in one beam size of 7 pixels for the off-position.}
\label{teles}
\end{figure*}

\begin{figure*}
\centering
\includegraphics[width=0.9\textwidth]{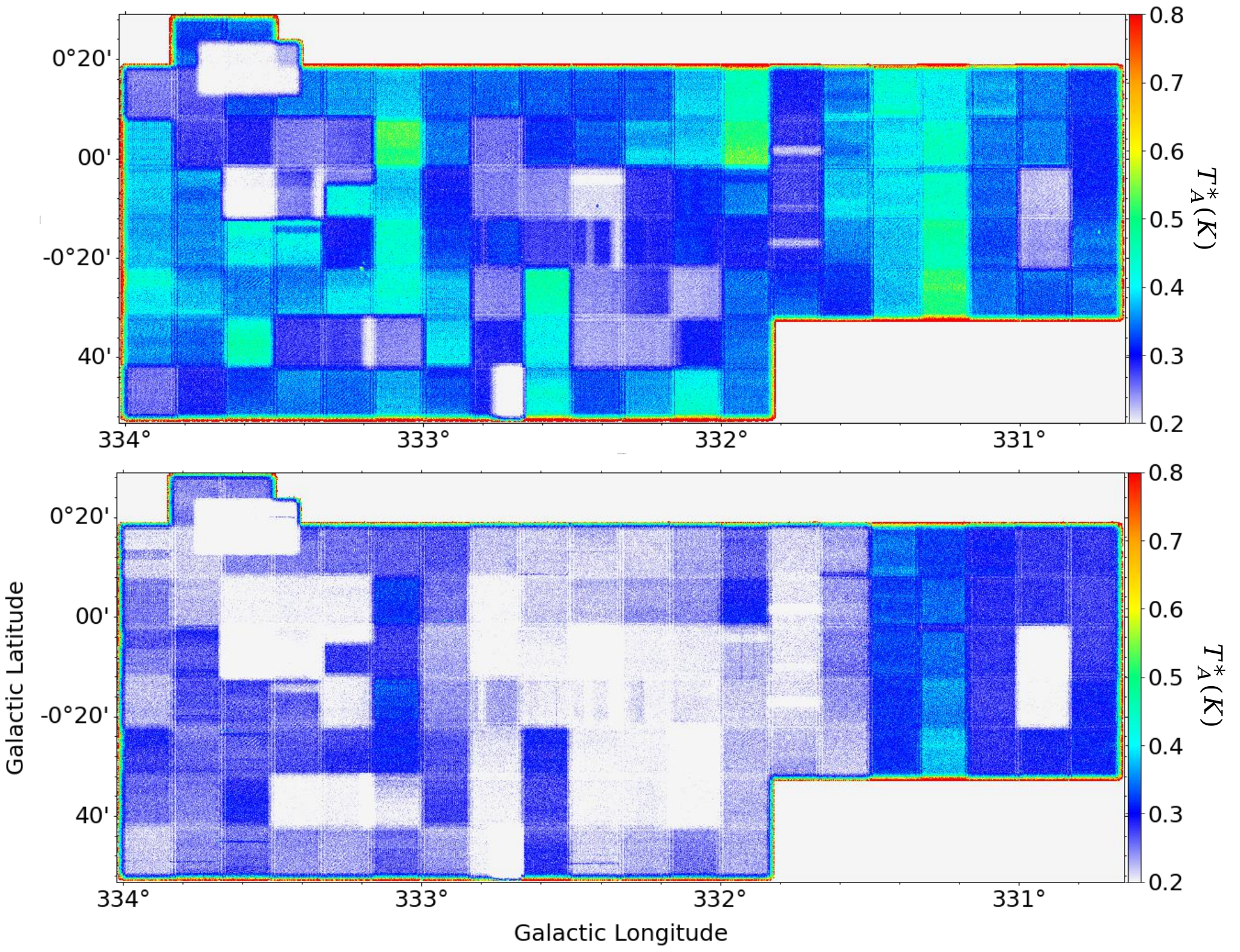}
\caption{Noise maps of $^{13}$CO (3-2) and $^{12}$CO (3-2) in our observation.}
\label{rms}
\end{figure*}

\begin{table}
\centering
\caption{Offsets of outer 6 pixels relative to the central pixel.}
\label{off}
\begin{tabular}{ccc}
\hline
pixel	&$\Delta$l (\arcsec)	&$\Delta$b (\arcsec)\\
	\hline
    1&0&0\\
    2&34.53 &-23.08 \\
    3&36.20 &19.15 \\
    4&1.74 &40.03 \\
    5&-35.78 &20.46 \\
    6&-35.33 &-19.86 \\
    7&-0.17 &-41.67 \\
	\end{tabular}
\end{table}

\begin{figure}
\centering
\includegraphics[width=0.45\textwidth]{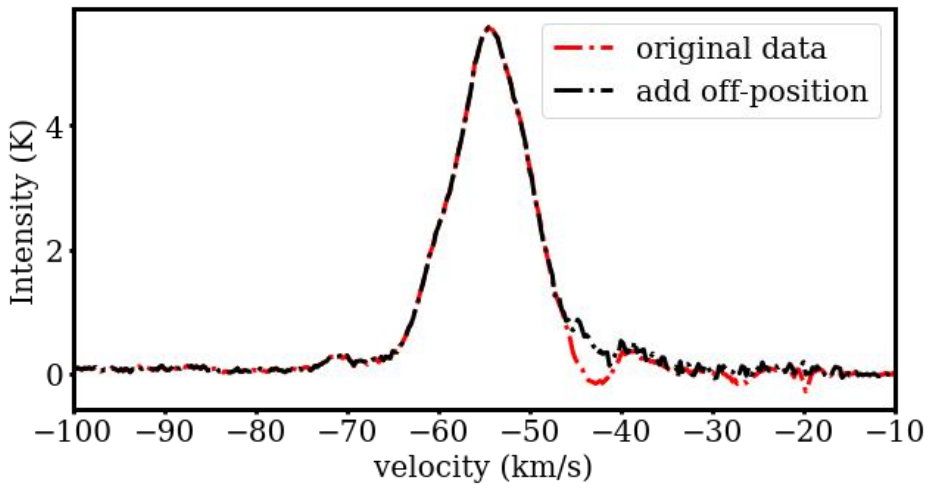}
\caption{Average spectra of one sub-map G333D55M-7M.}
\label{case}
\end{figure}

The receiver is a hexagonal array of six pixels surrounding a central pixel. The outer array is separated from the central pixel by $\sim 2$ FWHM. Table.\ref{off} lists the offsets of outer 6 pixels relative to the central pixel. Fig.\ref{off} shows the average spectra in one beam size of 7 pixels for the off-position. We can see the emission mainly concentrates on the velocity range [-50,-15] km s$^{-1}$, the absorption feature of one sub-map shown in Fig.\ref{case} also appears in the same velocity range, thus we add the emission of off-position in the velocity range [-50,-15] km s$^{-1}$ to each sub-map. As shown in Fig.\ref{case}, the absorption feature indeed disappears after adding the emission of off-position.

\end{appendix}

\end{document}